\documentclass[sigconf]{acmart}

\usepackage{etex}
\usepackage{balance}
\usepackage{graphicx}
\usepackage{hyperref}
\usepackage{multirow}
\usepackage[ruled,vlined,linesnumbered]{algorithm2e}
\usepackage{soul}

\usepackage{enumitem}
\usepackage{makecell}
\usepackage{pifont}

\def\Snospace~{\S{}}
\def\Fnospace~{\mbox{Fig.\hspace{0.25em}}}
\def\Tnospace~{\mbox{Tab.\hspace{0.25em}}}
\def\Enospace~{\mbox{Equation\hspace{0.25em}}}

\newcommand{\sysname}{Raven\xspace}

\newcommand{\companyname}{Microsoft\xspace}
\newcommand{\raven}{\sysname}
\newcommand{\mltosql}{MLtoSQL\xspace}
\newcommand{\mltodnn}{MLtoDNN\xspace}
\newcommand{\modelproj}{Model-projection pushdown\xspace}

\newcommand{\ie}{i.e.,\ }
\newcommand{\eg}{e.g.,\ }

\newcommand{\tinyskip}{\vspace{1pt}}
\newcommand{\mypar}[1]{\tinyskip\tinyskip\noindent\textbf{#1.}\xspace}

\newcommand{\myparrr}[1]{\tinyskip\tinyskip\noindent\underline{\em #1.}}
\newcommand{\mypari}[1]{\tinyskip\tinyskip\noindent\emph{#1}\xspace}

\newcommand{\vspaceafigure}{\vspace{-0mm}}

\definecolor{lightgray}{rgb}{.70,.70,.70}  
\definecolor{orange}{RGB}{255,127,0}
\sethlcolor{lightgray} 
\newcommand{\mynote}[1]{\par\noindent\colorbox{lightgray}{\parbox{\linewidth}{#1}}}
\newcommand{\mynotedark}[1]{\par\noindent\colorbox{black}{\parbox{\linewidth}{#1}}}
\newcommand{\mysimplenote}[1]{{#1}}

\newcommand{\KK}[1]{\textcolor{purple}{\mynote{KK:~#1}}}
\newcommand{\kk}[1]{\textcolor{purple}{\mysimplenote{[KK:]~#1}}}

\newcommand{\mi}[1]{\textcolor{brown}{\mysimplenote{[MI:]~#1}}}
\newcommand{\KP}[1]{\textcolor{red}{\mynote{KP:~#1}}}
\newcommand{\KS}[1]{\textcolor{cyan}{\mynotedark{KS:~#1}}}
\newcommand{\ks}[1]{\textcolor{cyan}{\mysimplenote{[KS:]~#1}}}
\newcommand{\kp}[1]{\textcolor{red}{\mysimplenote{[KP:]~#1}}}

\newcommand{\srev}[1]{#1}
\newcommand{\rev}[1]{#1}

\newcommand{\eat}[1]{}

\newcommand{\systext}[1]{\textsf{\small#1}}

\def\compactify{\noitemsep \itemsep=0pt \topsep=0pt \partopsep=0pt \parsep=0pt}
\let\latexusecounter=\usecounter

\newcommand{\mysim}{\raise.17ex\hbox{$\scriptstyle\sim$}}

\AtBeginDocument{%
  \providecommand\BibTeX{{%
    \normalfont B\kern-0.5em{\scshape i\kern-0.25em b}\kern-0.8em\TeX}}}

\copyrightyear{2022} 
\acmYear{2022} 
\setcopyright{acmlicensed}\acmConference[SIGMOD '22]{Proceedings of the 2022 International Conference on Management of Data}{June 12--17, 2022}{Philadelphia, PA, USA}
\acmBooktitle{Proceedings of the 2022 International Conference on Management of Data (SIGMOD '22), June 12--17, 2022, Philadelphia, PA, USA}
\acmPrice{15.00}
\acmDOI{10.1145/3514221.3526141}
\acmISBN{978-1-4503-9249-5/22/06}

\begin{document}
\fancyhead{}

\title{End-to-end Optimization of Machine Learning Prediction Queries}
\author{Kwanghyun Park, Karla Saur, Dalitso Banda, Rathijit Sen\\Matteo Interlandi, Konstantinos Karanasos}

\email{firstname.lastname@microsoft.com}

\affiliation{
  \institution{Microsoft}
}



\begin{abstract}


Prediction queries are widely used across industries to perform advanced analytics and draw insights from data. They include a data processing part (e.g., for joining, filtering, cleaning, featurizing the datasets) and a machine learning (ML) part invoking one or more trained models to perform predictions.
These parts have so far been optimized in isolation, leaving significant opportunities for optimization unexplored.



We present \sysname, \rev{a production-ready system for optimizing prediction queries.
\sysname follows the enterprise architectural trend of collocating data and ML runtimes.
It relies on a unified intermediate representation that captures both data and ML operators in a single graph structure to unlock two families of optimizations. 
First, it employs logical optimizations that pass information between the data part (and the properties of the underlying data) and the ML part to optimize each other.
Second, it introduces logical-to-physical transformations that allow operators to be executed on different runtimes (relational, ML, and DNN) and hardware (CPU, GPU).
Novel data-driven optimizations determine the runtime to be used for each part of the query to achieve optimal performance.}
Our evaluation shows that \sysname improves performance of prediction queries on Apache Spark and SQL Server by up to 13.1{\small$\times$} and 330{\small$\times$}, respectively.
For complex models where GPU acceleration is beneficial, \sysname provides up to 8$\times$ speedup compared to state-of-the-art systems.

\end{abstract}

\ccsdesc[500]{Information systems~Database query processing}
\ccsdesc[500]{Information systems~Query optimization}
\ccsdesc[500]{Computing methodologies~Machine learning}

\keywords{query processing, query optimization, machine learning prediction}

\maketitle

\section{Introduction}
\label{sec:intro}


Thanks to its recent advances, machine learning (ML) is being widely adopted in the enterprise and is on track to revolutionize every industry~\cite{ML-industry}. 
Trained ML models are being deployed in a wide variety of scenarios, ranging from datacenters to edge devices and across the software stack~\cite{cloud-to-edge}, to capitalize on the immense possibilities ML offers by drawing patterns and insights from data. 


High-value data in the enterprise is expressed, to a large extent, in structured or semi-structured forms~\cite{value-data,cloudy,no-left-behind}, typically stored in relational databases, data warehouses~\cite{polaris,redshift,snowflake}, or data lakes~\cite{adls,deltalake}. To train models on this data, data scientists often perform several pre-processing steps, e.g., to combine datasets (through unions, joins, aggregates), tokenize or unnest records, clean them, and encode them into features. 
Therefore, most widely used ML frameworks have added support for defining \textbf{trained pipelines}, i.e., Direct Acyclic Graphs (DAGs) of operators including ML models along with pre-processing steps~\cite{sparkmllib,sklearn,tfx,onnx,mlnet}.


\eat{
\begin{figure}[t!]
	\centering\includegraphics[width=\columnwidth]{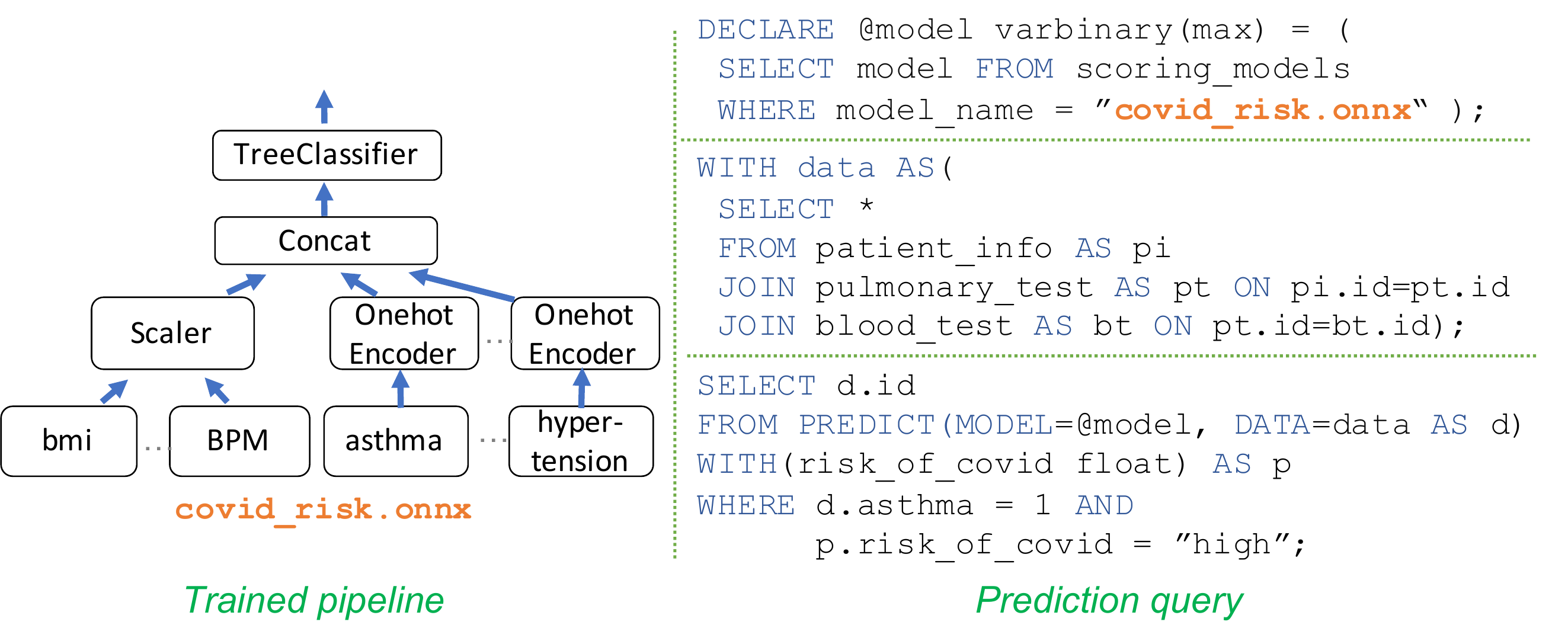}
 	\vspace{-7mm}
	\caption{Prediction query using SQL Server's \textsc{predict} statement (right) to invoke a trained pipeline (left). The query includes data processing operations (e.g., join, one-hot encoding) and a tree-classifier model.}
    \label{fig:run-example}
    \vspace{-5mm}
\end{figure}
}

\mypar{ML inference in the enterprise}\footnote{Hereafter, we use the terms prediction and inference interchangeably.} Data analysts and software engineers issue \textbf{prediction queries}, \rev{which are complex analytics queries} that employ trained pipelines to perform predictions over new data arriving in the database/data lake. These queries often include additional data processing operators (e.g., filters or joins), implementing prediction-specific logic. 
\rev{As we discuss in \autoref{sec:archi}, ML inference drives the majority of the cost (up to 90\%~\cite{reinvent2018}) associated with ML in the enterprise, whereas traditional ML models (i.e., non-neural networks, such as linear regression and tree-based models) are still the most widely used by a large margin~\cite{kagglesurvey,dsonds}. Moreover, based on the analysis of $130$ customer engagements at \companyname, we found that batch inference is chosen over online inference in most enterprise scenarios.}
\rev{Not surprisingly}, most cloud vendors have enabled issuing batch prediction queries through recent SQL extensions or traditional user-defined functions (UDFs) for inference~\cite{predictsql,sparkpredict,big-query-ml,redshift-ml,sparksql,madlib}.
\rev{Therefore, in this work we focus on the optimization of batch prediction queries that invoke traditional ML models.}



\mypar{Optimization opportunities}
\rev{We make two crucial observations that open untapped opportunities for the optimization of prediction queries.}
First, existing work has mainly focused on optimizing each part of the prediction queries in isolation~\cite{tvm,pretzel,willump}.
However, at their core, prediction queries are just a set of operations over data, which, treated holistically, offer great opportunities for optimization---we provide an analysis of hundreds of pipelines in \autoref{sec:archi}.
\srev{With Raven's initial version~\cite{raven}, we laid out our vision of optimizations spanning data and ML operators in prediction queries. However, in that initial prototype, a few limited optimizations were applied manually over simplistic single-operator models---our analysis in \autoref{sec:archi} shows that such models are uncommon in practice.}
Recently, a few more works have started touching upon optimization across data and ML operators, but only in the context of training and generalized linear models~\cite{systemml,lara,spores,learning-joins,relational-borg} or targeting their execution at the physical level~\cite{weld}.


\rev{Second, most related research efforts consider a unified runtime to execute relational and ML queries~\cite{systemml,weld,lara}. In industry-grade data systems, however, we observe a different trend: while some of these ideas can be applied on parts of the query, production systems have steered away from running every ML workload natively. This is particularly true for complex data operations (e.g., tf.data allows only simple transformations~\cite{tf-data}) or complex ML operations (linear algebra may be efficiently executed inside the database but covers small portion of prediction pipelines). Instead, modern data engines integrate with ML runtimes for breadth of ML support: Microsoft SQL Server/ONNX Runtime~\cite{predictsql}, Google BigQuery/TensorFlow~\cite{big-query-ml},  Amazon Redshift/SageMaker~\cite{redshift-ml}.}


\rev{We believe that the co-existence of specialized (software/hardware) accelerators into data runtimes is not the outcome of current technology limitations but an architecture that is here to stay. Thus, the natural question to ask is: having these ML runtimes next to the data engine, how and when should we exploit each runtime to accelerate prediction queries?} 

\mypar{\sysname Optimizer} \srev{Through this work, we substantially push the envelope in realizing and extending our \sysname vision~\cite{raven} for the optimization of prediction queries.
Following the architecture of collocating data engines with ML accelerators, we present the optimizer that fuels \sysname, which is the result of 10 people-years of work. Our optimizer (i)~applies logical optimizations to holistically optimize the query; and (ii)~judiciously picks which part of the plan to run on each engine.} Concretely, we make the following contributions.

\myparrr{State of enterprise ML inference (\autoref{sec:archi})}
\rev{We provide several insights for the state of ML inference in the enterprise, based on findings from \companyname, customer engagements, offerings from various cloud vendors, and an analysis of hundreds of publicly available pipelines. All these acted as a motivation for building \sysname.}

\myparrr{Logical optimizations (\autoref{sec:opt})}
\sysname relies on a unified intermediate representation (IR) that captures both data and ML operators in a common \rev{structure}~(\autoref{sec:ir}). It \rev{combines} concepts from the relational algebra and similar efforts recently introduced \rev{by ML frameworks (namely, ONNX~\cite{onnx})}. 
Having all operators of a prediction query in a single IR allows us to optimize the query holistically. 
\rev{In particular, we extend the \emph{cross-optimizations} (first introduced in~\cite{raven}) to support a wide range of models with multiple operators each, which are actually used in the enterprise (including several tree-based models, such as random forests and gradient boosting trees). We also introduce novel \emph{data-induced} optimizations.
Through these logical optimizations, \sysname exploits features of data processing operators and data properties to avoid unnecessary computation in the ML part and vice-versa by flowing information between operators~\cite{sipRaghu}. For example, a data predicate on an input feature can be used to simplify the model at compile time, whereas input features that end up not participating in inference can be completely removed from the whole query.}



\myparrr{Logical-to-physical: runtime selection (\autoref{sec:opt:logtophys})}
\rev{The collocation of the data engine with ML accelerators allows \sysname to assign different parts of the IR to the runtime that will lead to optimal performance. To this end, we employ logical-to-physical optimizations that can turn a classical ML model to an equivalent SQL statement (to be executed by the data engine) or to an equivalent neural network (to be executed by the ML accelerator, possibly on a GPU).}

\rev{Most importantly, we show that these optimizations are not always beneficial: blindly applying them, \srev{as we did in the initial \sysname prototype~\cite{raven}}, can lead to slowdowns of up to 6{\small $\times$}. Hence, \sysname introduces optimization strategies to pick the rules that would lead to best performance for each query on a given hardware setup. 
Instead of relying on hard-coded heuristics that do not work across workloads and hardware, \sysname employs data-driven, instance-optimized strategies: a novel \emph{data-informed} strategy and two ML-based ones.}




\myparrr{An end-to-end system (\autoref{sec:impl})}
\rev{The \sysname optimizer, which has been fully implemented as an extension to Apache Spark, is the first production-ready system to incorporate such optimizations. Users specify prediction queries in SparkSQL invoking models expressed in ONNX (or any format convertible to ONNX~\cite{onnxconvert}). \sysname's optimizer with all the above optimizations and optimization strategies are implemented as a co-optimizer invoked by Spark's Catalyst optimizer.
We have also made it possible to execute \sysname's optimized plans in SQL Server.}



\vspace{1mm}
We experimentally validate the benefits of \sysname over a variety of datasets, prediction queries, and hardware (\autoref{sec:eval}). 
Our results show significant performance gains on both Spark and SQL Server. In particular, \sysname's optimized plans achieve 1.4--13.1{\small$\times$} speedup against Spark and 1.4--330x against SQL Server. We also report similar and even bigger gains against other state-of-the-art systems, \rev{such as MADlib~\cite{madlib}}. Moreover, we show that complex models benefit from GPU acceleration by up to 8{\small$\times$}.

\sysname's execution of prediction queries, \srev{using the newly introduced {\sc predict} statement, became publicly available recently as part of \companyname's Spark cloud offering~\cite{sparkpredict}}. We are currently working closely with the product groups to make the optimizer also available. 
\sysname is only the first click-stop in our long-term vision---we are exploring, for example, when and how to offload relational and graph operators to ML accelerators \srev{with very promising initial results~\cite{surakav-vision}}.

\vspace*{-1mm}
\section{Overview}
\label{sec:archi}

In this section, we provide evidence data about the state of ML in the enterprise, which motivated us to focus on the problem of optimizing batch prediction queries that involve traditional ML models. Then we give an overview of \sysname.


\begin{figure}[t]
	\centering\includegraphics[width=1.0\columnwidth]{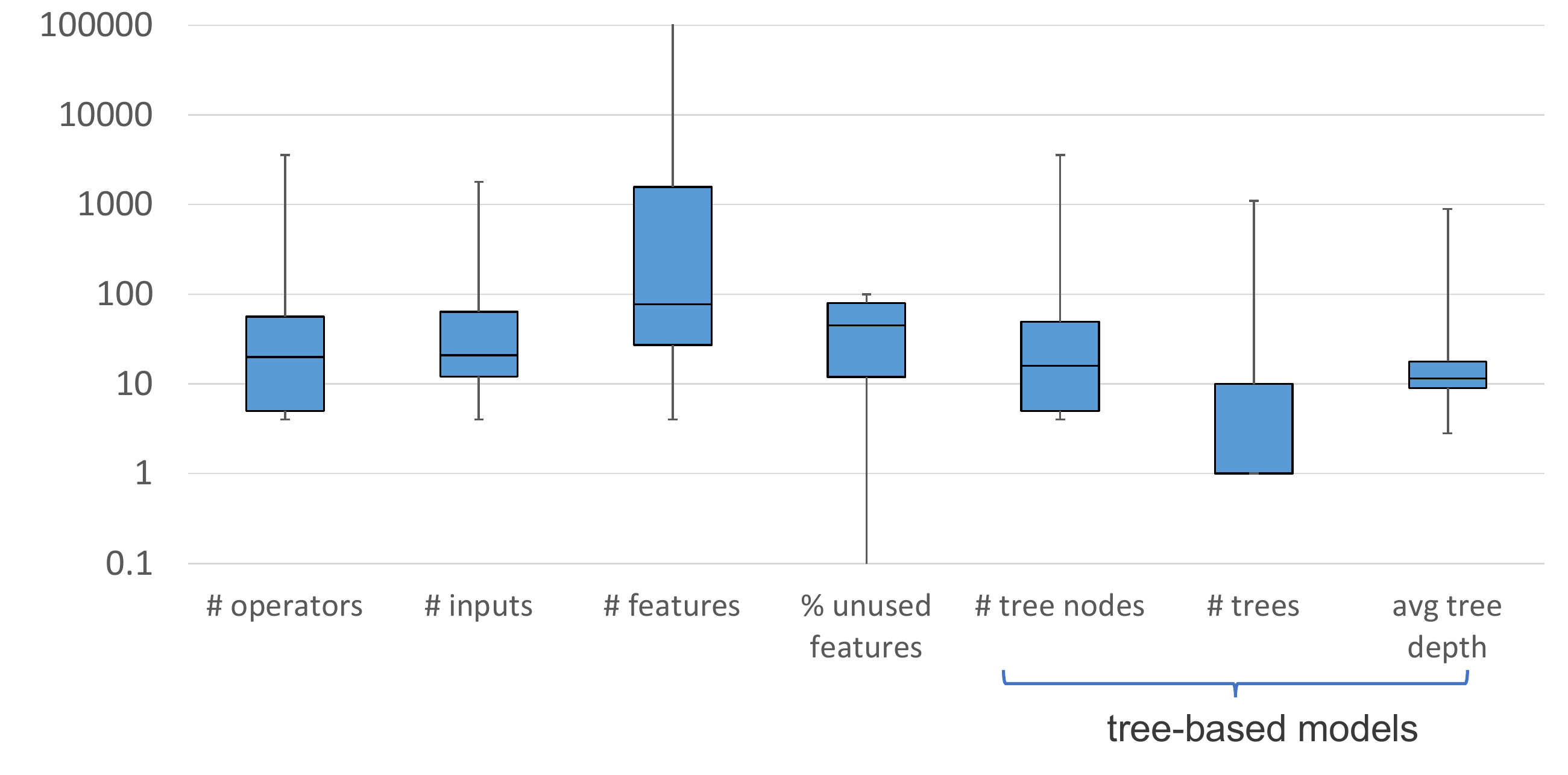}
	\vspace{-8mm}
	\caption{Statistics for \texttildelow$500$ traditional ML models in the OpenML CC-18 benchmark~\cite{openml_cc18}. The boxplots represent the 25$^{\textrm{th}}$, median, and 75$^{\textrm{th}}$ percentile; the whiskers represent the min and max. Y-axis is in logscale.}
    \label{fig:openml-stats} 
    \vspace{-7mm}
\end{figure}

\subsection{Motivation}
\label{sec:over:motiv}




\mypar{Inference drives the cost of ML in the enterprise} 
In most applications, models are trained at regular intervals but used for inference continuously. Although a training instance might be more costly than an inference one, inference ends up being significantly more costly than training on aggregate. Cloud vendors state that 90\% of the total cost of ML is on inference~\cite{reinvent2018}. Therefore, optimizing inference is crucial for lowering operational costs.

\mypar{Batch inference is often preferable or at least sufficient} Online inference is not a requirement for most enterprise applications, especially after considering the infrastructure costs associated with it.
In $130$ customer engagements at \companyname, the requirements of $91\%$ of them were captured by batch inference. 
An additional $6\%$ (for a total of $97\%$) was managed with batch inference at short intervals. Hence our focus with \sysname is on batch prediction queries.

\begin{figure}[t!]
	\centering\includegraphics[width=\columnwidth]{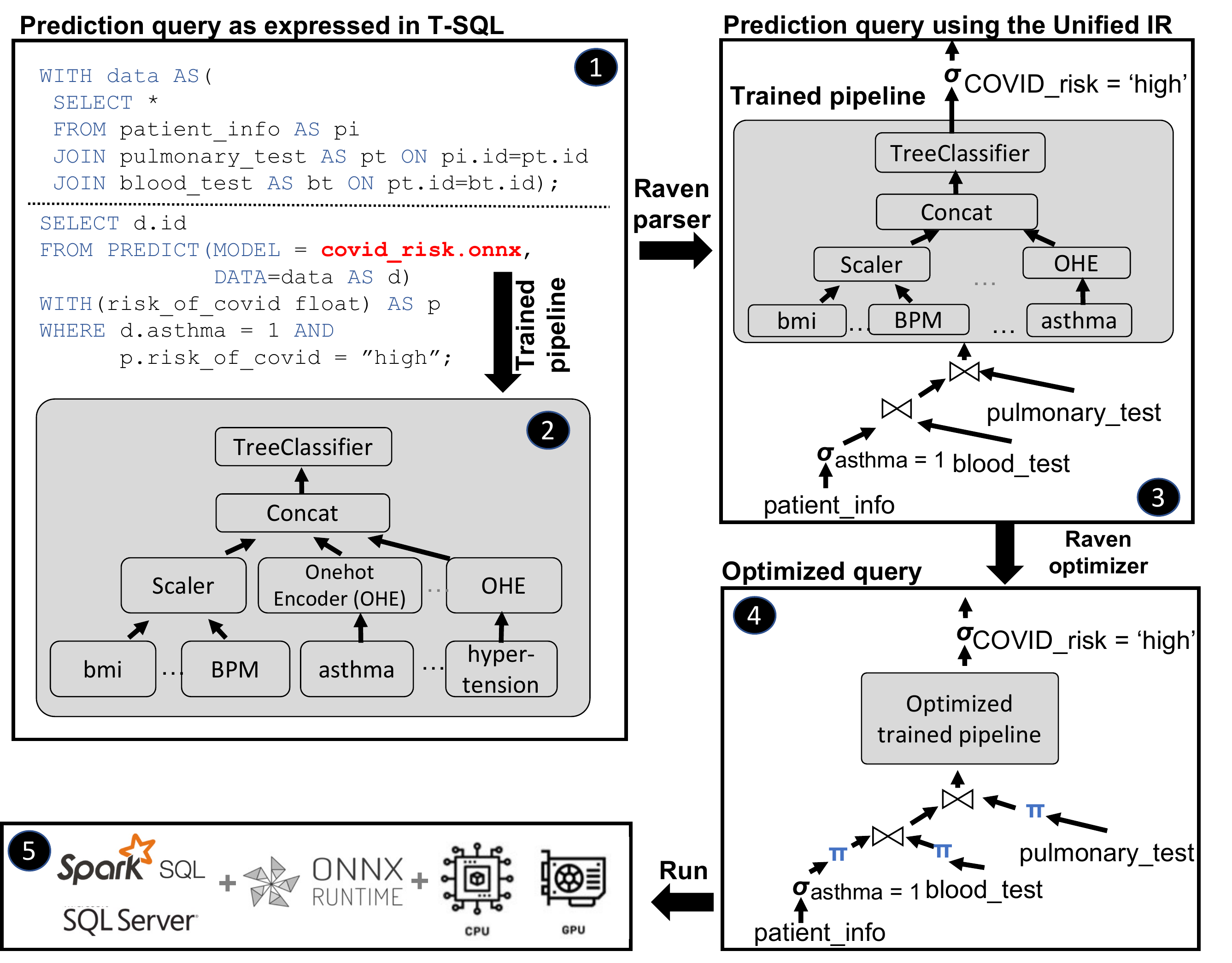}
	\vspace{-7mm}
	\caption{Prediction query using SQL Server's \textsc{predict} statement \ding{202}. It includes data processing operations (e.g., join, one-hot encoding) and invokes a trained pipeline \ding{203}. \sysname: \ding{204} constructs the corresponding intermediate representation (IR); \ding{205} optimizes the IR through cross-optimizations and transformations; and \ding{206} executes the final optimized plan using Apache Spark or SQL Server (depending on where the data is) and an ML runtime (if needed) over CPU or GPU.}
    \label{fig:ravenflow}
    \vspace{-6mm}
\end{figure}

\mypar{Prediction queries in SQL}
Input to \sysname are prediction queries, i.e., advanced analytics queries that process data residing in (local or remote) files or databases through various data transformation operations and feed them into one or more trained pipelines.
Two main ways are used to express such queries: SQL and Python. In this work, we consider the SQL syntax as adopted by SQL Server offerings (including Azure SQL DB and DW~\cite{predictsql}), i.e., a {\sc predict} table-valued function (TVF) accepting as parameters a model (the trained pipeline) and a table, as depicted in \autoref{fig:ravenflow} (\ding{202}). Similar SQL-based syntax is used by both Google BigQueryML~\cite{big-query-ml} and Amazon RedshiftML~\cite{redshift-ml}, and recent work can be used to translate prediction queries from Python to SQL~\cite{magpy}.

\mypar{Traditional ML is most widely used}
According to the latest Kaggle survey~\cite{kagglesurvey} and an analysis of publicly available Python notebooks~\cite{dsonds}, traditional ML algorithms, such as linear/logistic regression and tree-based models (decision trees, random forests, gradient boosting) are the most popular by a large margin. \texttildelow$80\%$ of the Kaggle responders use them, as opposed to $43\%$ for neural networks. Scikit-learn~\cite{sklearn-paper}, which focuses on traditional ML, is the most widely used ML library in both studies. Thus, while \sysname can execute queries including any model expressible in ONNX~\cite{onnx}, our current focus is on optimizing queries with traditional ML operators.

\mypar{Trained pipelines \rev{are complex and} vary greatly}
We studied $508$ scikit-learn trained pipelines over $72$ datasets from OpenML's CC18 benchmark suite~\cite{openml}. 
\autoref{fig:openml-stats} highlights the significant variations across pipelines. Pipelines have a median of $21$ inputs but some receive more than $1000$ inputs. After featurization, due to categorical inputs, models have a median of $77$ features but some models have more than $50M$ features. Likewise, while most tree-based models (which account for $88\%$ of all models) include less than $10$ trees with a median depth of $11$, some are extremely complex with thousands of trees and depth. \rev{Each model consists of a large number of operators with an average of $126$, a median of $20$, and a min/max of $4$/$3560$, as shown in \autoref{fig:openml-stats}}. Similar model complexity variations have been reported in production settings~\cite{mlnet}. 
These variations make different transformations be beneficial for each model and dictate data-driven optimization strategies to determine which rules to apply (\autoref{sec:opt:logtophys}).

Finally, to avoid overfitting in linear models, regularization techniques (e.g., Lasso)~\cite{bishop-book} are often used during model training, which end up creating zero weights. In tree-based models, features are often left out during training due to correlations or insignificant contributions. 
In the $508$ OpenML models we analyzed, on average $46\%$ of the model features remain unused during inference!
This observation makes the model-projection pushdown rule particularly effective, as we discuss in \autoref{sec:opt:crossir}.




\subsection{\sysname}

\mypar{Running example} Consider a model that \emph{predicts whether a patient is in high risk of COVID-19 complications}. 
This model is trained over a large amount of data across several COVID-19 testing sites and hospitals. The result of model training is the model pipeline $M$, shown in \autoref{fig:ravenflow} (\ding{203}). Along with the actual tree classifier, it includes pre-processing steps to prepare the data for inference. $M$ is deployed, and then a data analyst can use it in a prediction query $P$ to ``{\em find asthma patients who are likely in the high-risk COVID-19 group},'' depicted in \autoref{fig:ravenflow} (\ding{202}). 
The analyst uses $P$ to find such patients in a specific hospital and to do so, $P$ first joins \verb|patient_info| with two test results (\verb|blood_test|, \verb|pulmonary_test|) and then invokes $M$. $P$ also includes filters on \verb|asthma| and the result of $M$.

\mypar{Building the unified IR}
Given a prediction query $P$, which includes both data processing operators in SQL and the trained pipeline $M$, \sysname parses $P$ and constructs a DAG expressed in \sysname's unified IR (\autoref{fig:ravenflow} (\ding{204})). 
The unified IR contains a mix of ML operators (the {\sf TreeClassifier}, {\sf Scaler}, and {\sf OneHotEncoder}) and relational ones (e.g., the selection predicate over \texttt{asthma} and the joins between the three tables).
The unified IR unlocks \rev{several logical and logical-to-physical optimizations}.







\mypar{Optimizations}
\sysname's optimizer is a co-optimizer sitting alongside the optimizers of the data and ML engine, and is invoked before those optimizers. 
Given the IR corresponding to an inference query $P$, including both data and ML operators, the \sysname optimizer triggers several optimization rules in the form of passes over the IR (\autoref{fig:ravenflow} (\ding{205})), \rev{falling in two main categories:}

\mypari{Logical optimizations} \rev{that simplify the data part of the query by taking into account information from the ML part and vice versa (cross-optimizations) or that use data statistics to simplify the query (data-induced). Optimizations in this category are always beneficial and are applied first (similar to the heuristic optimizations always applied first by relational optimizers).}

\mypari{Logical-to-physical optimizations} \rev{that transform traditional ML operators into equivalent SQL statements or neural networks to be executed by the data engine or the neural network runtime (possibly over GPU), respectively.
\sysname uses data-driven optimization strategies for determining which transformation to apply given $P$.}

\noindent Note that well known optimizations are also triggered by the data and ML engines that \sysname uses prior to executing $P$ (e.g., projection or aggregate pushdowns, join eliminations, compiler optimizations).

\mypar{Execution of optimized plan}
The final optimized prediction query will be executed by the data engine that owns the data (\autoref{fig:ravenflow} (\ding{206}))---Apache Spark or SQL Server in our implementation. The data engine will perform all data operations (\ie SQL operators, as well as ML operators that got translated to SQL by \sysname) and will invoke ML runtimes as required for evaluating the trained pipelines.
In particular, ONNX Runtime~\cite{ort}  
is integrated with SQL Server~\cite{raven,predictsql} and can be invoked through the {\sc predict} statement. We added a similar syntax in Apache Spark to invoke ONNX Runtime (both on CPU and GPU). More implementation details are discussed in \autoref{sec:impl}.

\vspace*{-1mm}
\section{\sysname's Unified IR}
\label{sec:ir}


IRs have been commonly used for optimizations in various settings. Most database optimizers rely on relational algebra~\cite{cowbook}, whereas different IRs are used in ML runtimes (e.g.,~\cite{xla,onnx,mlir,tvm,halide}) \rev{or to combine relational and linear algebra operators for ML training~\cite{lara,laradb}.
In \sysname, we represent a wider range of data and ML operators in a unified IR, which allows us to perform logical optimizations holistically (\autoref{sec:opt}) and perform runtime selection (\autoref{sec:opt:logtophys}). 
Our IR's operators fall in the following categories:}
\begin{itemize}[noitemsep,topsep=2pt,parsep=2pt,partopsep=2pt,leftmargin=10pt]

\item \textbf{Relational algebra.} This includes all the relational algebra operators, which are found in a typical RDBMS.

\item \textbf{Linear algebra.} A large fraction of the operators used in ML frameworks, and in particular neural network runtimes~\cite{tensorflow,ort,pytorch}, fall into this category. Examples include {\sf matrix multiplication} and {\sf convolution} operators.

\item \textbf{Other ML operators and data featurizers.} These are all the operators widely used in traditional ML frameworks (e.g., scikit-learn~\cite{sklearn}, ML.NET~\cite{mlnet}) but cannot be expressed in linear algebra. Examples are
decision trees and featurization operations (\eg categorical encoding, text featurization). \rev{Note that supporting such operators is crucial (see \autoref{sec:over:motiv}) but also challenging, as they do not abide by an algebra and can correspond to arbitrary algorithms.}
\end{itemize}

\vspace{1mm}
\srev{The main motivation when designing our IR was to (i)~ease production adoption, and (ii)~be able to express a wide range of prediction queries. To this end, we made the following design choices:}
\begin{itemize}[noitemsep,topsep=2pt,parsep=2pt,partopsep=2pt,leftmargin=10pt]
\item \srev{We based our IR implementation on ONNX, which we extended with common relational operators (e.g., projections, selections), rather than start from a relational IR. This was a pragmatic choice as there are many more ML operators than relational ones.}
\item \srev{We chose ONNX for the ML part, as most popular frameworks translate to it. Our current set of ML operators is all ONNX operators, including featurizers (e.g., one-hot encoder, scaler, normalizer, etc.), linear algebra, and traditional ML ones (e.g., decision tree, random forest, gradient boosting, linear regression).}
\item \srev{Unlike most systems, we support more operators than just linear algebra for ML. Linear algebra is easily executed in the DB but lacks expressivity for models used extensively in production, e.g., featurizers and tree-based models.} 
\end{itemize}

\vspace{1mm}
\srev{Embracing widely used standards, namely SQL and ONNX, in our IR was key for production adoption. For example, on the relational side, all standard SQL operators in SQL Server and SparkSQL can be mapped to \sysname IR operators. Similarly, on the ML side, by supporting ONNX, \sysname also supports by transitivity all models that can be converted to ONNX~\cite{onnxconvert}, such as scikit-learn, SparkML, PyTorch, and TensorFlow.}
Given there is an 1-1 mapping between SQL/ONNX operators and our IR operators, it is also straightforward to construct the IR by traversing the prediction query.

Note that some operators are semantically very close, e.g., {\sf \small FeatureExtractor} (common in ONNX graphs) and the relational Projection---we allow both in our IR and have equivalence rules that let us translate from one to the other. Models with operators that are not yet supported are executed but not optimized by \sysname.

\section{Logical Optimizations}
\label{sec:opt}

\begin{figure*}[]
\vspace{-2ex}
\hspace{-3ex}
{\includegraphics[width=\textwidth]{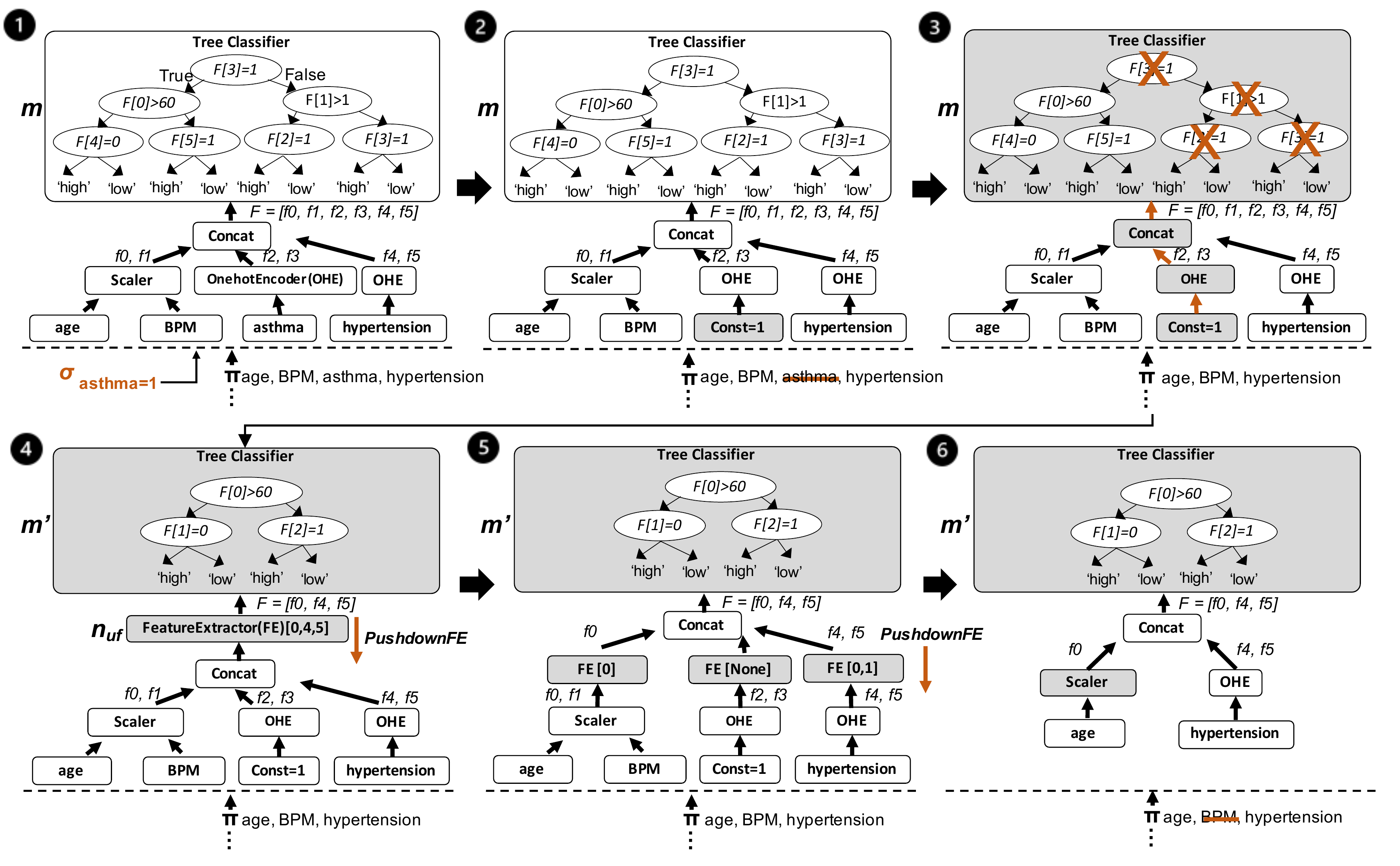}}
\vspace{-5mm}
\caption{Cross-optimization steps for our running example (we omit some data operators from the IR): predicate-based model pruning (\ding{202}--\ding{204}) and model-projection pushdown (\ding{205}--\ding{207}). We use gray and orange to highlight the changes in the IR at each step.}
\label{fig:cross_ir_example}
\vspace{-5mm}
\end{figure*}


\rev{In this section, we describe two types of logical optimizations applied by \sysname's optimizer: (i)~cross-optimizations (\autoref{sec:opt:crossir}) that exchange information between the data and ML parts of the query to optimize each other; and (ii)~data-induced optimizations (\autoref{sec:opt:datainduced}) that exploit statistics of the underlying data to further simplify the query plan. 
All these are implemented as rules over \sysname's IR.\footnote{We use prediction queries and their corresponding IR interchangeably in this section.}}




\subsection{Cross-optimizations}
\label{sec:opt:crossir}


\eat{
\begin{table}[ht]
\caption{Notation used in Section \autoref{sec:opt:crossir}} 
\centering 
\begin{tabular}{c l} 
\hline\hline 
Symbol & Description \\ [0.5ex] 
\hline 
${N}$ & List with all \sysname IR nodes\\
$P$ & List of data predicates\\
$N.attr$ & Attribute \emph{attr} of \sysname IR node\\
\hline 
\end{tabular}
\label{table:notation} 
\end{table}
}

In these optimizations, we leverage properties of the data processing part of the prediction query to optimize the ML part (\emph{data-to-model}) and vice-versa (\emph{model-to-data}). These optimizations can be seen as similar in spirit to sideways information passing at compile time~\cite{sipRaghu}. Below we describe one optimization from each category.
%
\autoref{fig:cross_ir_example} shows several versions of the IR when applying the two optimizations on the prediction query of \autoref{fig:ravenflow} (\ding{202}).

\srev{We first introduced the two optimizations below in~\cite{raven}, supporting only models with a single linear regression or decision tree operator.}
We are expanding these optimizations here in two main ways: (i)~we support a much wider range of traditional ML operators, including all common linear and tree-based models (decision trees, random forests, gradient boosting) and featurizers, covering all operators present in the hundreds of OpenML pipelines we studied in \autoref{sec:archi} \srev{(for comparison, none of the OpenML operators were supported in~\cite{raven})}; and (ii)~we support arbitrary ONNX models instead of single operators. \srev{A manual approach as in~\cite{raven} could not handle this level of complexity: an actual optimizer is required to traverse the operator tree and apply the optimizations, propagating them through featurizers and other operators.}

\mypar{Predicate-based model pruning}
Given a prediction query $P$, this data-to-model optimization identifies data predicates in $P$ (from the {\sc where} clause of the query) and passes them to the trained pipelines in $P$ to simplify them, resulting in an optimized prediction query $P'$.
The benefit of this optimization is twofold: (i)~it may reduce the number of inputs to the trained pipeline (if an input is a known constant, we do not need to provide it to trained pipeline at runtime); and (ii)~it reduces the complexity of the models, e.g., by pruning tree-based models or statically pre-computing parts of the ML operations (e.g., multiplications in linear models).

For each trained pipeline $M$ of the query, the algorithm proceeds in two steps:

\noindent\emph{Step 1.} It collects $M$'s inputs that participate in predicates. For each equality predicate, it replaces the corresponding model input with a constant node and adds a projection to prevent that input from reaching the model. This projection can then be pushed down by the relational optimizer to completely avoid scanning that column---with sufficient predicates even full table scans can be avoided.
It also annotates with predicate information the model inputs corresponding to range predicates.
In \autoref{fig:cross_ir_example}~(\ding{203}), given \verb|asthma=1|, the algorithm replaces the \verb|asthma| input with a constant node and removes the corresponding data column from the {\sf project} operator.

\vspace{1mm}
\noindent\emph{Step 2.} It passes the equality/range predicate information through the pre-processing/featurization operators of $M$ (\eg {\sf Scaler}, {\sf OneHotEncoder}, {\sf Concat}), updating the predicate information as needed. Once it reaches a tree-based model (\eg decision tree, random forest, gradient boosting), it uses this information to prune branches of each tree in the model.

In \autoref{fig:cross_ir_example}~(\ding{204}), predicate \verb|asthma=1| becomes $[0,1]$ when pushed through the {\sf OneHotEncoder} for the two categories [is\_not\_asthma\_patient, is\_asthma\_patient]. Similarly, a constant $n_c$ is updated to $(constant - offset) \times scale$ when pushed through a {\sf Scaler}.
Moreover, {\sf TreeClassifier}'s root and right branch were completely pruned based on \verb|asthma=1|. It can also leverage range predicates, e.g., with \verb|age<30|, it can further prune the right branch of the tree in \autoref{fig:cross_ir_example}~(\ding{205}).




Predicate-based model pruning can be particularly effective with predicates over a categorical input $c_i$ with $V$ unique values. These inputs are typically one-hot encoded to $V$ binary variables, and an equality predicate on $c_i$ leads to $V$ known constant model inputs.

We also support predicates on the outputs of the trained pipelines, such as \verb|risk_of_covid='high'| in \autoref{fig:ravenflow}. In the case of tree-based models, we pick the leaves that satisfy the predicate and traverse the model bottom up, keeping the paths from these leaves to the model inputs and pruning all other nodes.

\vspace{1mm}
\mypar{Model-projection pushdown}
This model-to-data optimization relies on the model sparsity observation we made in \autoref{sec:over:motiv} about features often left unused during inference.
Features might also be deemed unnecessary due to other optimizations: after predicate-based model pruning, the features of the right tree branch in \autoref{fig:cross_ir_example}~(\ding{204}) are no longer used.
Model-projection pushdown identifies such unused features and attempts to push them down below the trained pipeline, towards the data processing part. This can bring significant performance benefits: (i)~if we push the projection all the way to the scan, we completely avoid reading some features from disk; (ii)~even if this is not possible, we might still be able to push them below joins, thereby significantly reducing shuffling costs, or avoid those joins altogether if all features of a table end up getting projected out.
%
%
%

\rev{This optimization is implemented as two passes over the IR, each applying a transformation rule until fixpoint.
In the first pass,} for each model $m$ in the trained pipeline $M$, it detects its unused features and creates a dense version $m'$ of it.\footnote{We densify a linear/logistic regression by removing zero-weight features, and a tree-based model by removing the inputs not used by any tree.} It then replaces $m$ with $m'$ and projects out the unused features by adding a FeatureExtractor node $n_{uf}$ in $M$.\footnote{FeatureExtractors are similar to relational projections, used in ML frameworks.}
\rev{In the second pass, a rule is triggered on} every newly added $n_{uf}$ to push it down to its parents in $M$.


In \autoref{fig:cross_ir_example}~(\ding{204}), the original {\sf TreeClassifier} using only the features with indices [0,4,5] is replaced by the densified model $m'$ in \autoref{fig:cross_ir_example}~(\ding{205}) using indices [0,1,2] along with a FeatureExtractor node $n_{uf}$ to project out unused features from the model.
In \autoref{fig:cross_ir_example}~(\ding{206}), the {\sf FeatureExtractor} is pushed below {\sf Concat} operator, creating three {\sf FeatureExtractors}. The middle one is empty and is removed from the DAG along with its parents. The left-most is further pushed below {\sf Scaler}, and the unused input \verb|BPM| is removed from the pipeline and from the list of input columns in the {\sf Project} operator (\autoref{fig:cross_ir_example}~(\ding{207})).

\vspace{1mm}
\mypar{Extension to neural networks}
\srev{Based on the evidence data we presented in \autoref{sec:over:motiv}, we have focused our efforts so far on traditional ML operators, such as linear and tree-based models and featurizers (see \autoref{sec:ir}), given their popularity. Thus, our optimizations cannot be applied to neural networks yet. Predicate-based model pruning in a neural network would amount to constant folding. Model-projection pushdown might be less beneficial on fully connected models like MLPs, but it remains open for sparser models.}

\subsection{Data-induced Optimizations}
\label{sec:opt:datainduced}


\rev{
RDBMS as well as big data systems, such as Apache Spark, commonly store data statistics, such as maximum and minimum values for each data column, to allow the optimizer to generate better, more accurate query plans. 
Data statistics can also be used in concert with data partitioning to further speed up query execution, for instance by means of data skipping~\cite{data-skipping}.}

\rev{
In \sysname, we make use of data statistics to optimize the ML models of a prediction query. For example, \texttt{age} is the root feature of the tree of \autoref{fig:cross_ir_example}~(\ding{207}), and any instance with \texttt{age} $>$ 60 will go in the left sub-tree of the tree, or right otherwise. However, if the input data contain no instance with \texttt{age} $\leq$ 60, the tree can be simplified by completely pruning the right sub-tree, also removing the participating data inputs (\texttt{age} in our example).
\sysname induces such predicates based on min/max data statistics.
}


\rev{Taking this idea a step further, we also exploit the fact that big data systems store data in partitions~\cite{partition-discovery}, which are either specified by the user or are the result of a data operation (e.g., group-by).
Following this intuition, \sysname compiles an optimized model for each partition, leveraging the data distribution of that partition.
Going back to our previous example, if data is partitioned and distributed over the \texttt{age} feature, \sysname can generate partition-optimized trees by pruning the left or the right sub-tree based on the maximum and minimum value contained in each partition.
Note that data-induced predicates can work with the other cross-optimizations: i.e., if a feature is pruned from a model thanks to a data-induced predicate, the model-projection pushdown optimization can further prune the column from the inputs if not used elsewhere.
}

\vspace*{-1mm}
\section{Logical to Physical: Runtime Selection}
\label{sec:opt:logtophys}


\rev{We now focus on logical-to-physical optimizations that aim to convert (a set of) operator(s) to another in order to allow more efficient engines and hardware to be used for executing those operators~(\autoref{sec:opt:optransform}).
A crucial observation is that such rules can be detrimental if not applied with care. For example, in our experiments, turning a scikit-learn gradient boosting classifier with default values ($100$ estimators, max-depth $3$) to a SQL statement led to a 6{\small $\times$} slowdown. Therefore, we introduce three optimization strategies to judiciously pick when to apply each transformation (\autoref{sec:opt:opt}).}

\subsection{Transformations for Runtime Selection}
\label{sec:opt:optransform}

\mltosql turns ML operators to SQL statements to avoid invoking the ML runtime for those operators, whereas \mltodnn transforms traditional ML operators to equivalent (deep) neural networks (DNN) to make use of state-of-the-art DNN runtimes and compilers over modern hardware.





\mypar{\mltosql}
This optimization converts ML operators (linear algebra and other ML operators and featurizers; see \autoref{sec:ir}) to equivalent (\ie semantics-preserving) relational ones. The main benefit of this conversion is to reduce or completely avoid (when we can convert the whole model pipeline to SQL) invoking the ML runtime, thereby avoiding initialization costs and data conversions/copies between the relational and ML engines. It can also enable more extensive relational optimizations in the DBMS, \eg pushing ML computation (now expressed as relational operators) below joins and aggregates.


\sysname supports the conversion of a wide range of traditional ML models to SQL, such as linear models (\eg logistic/linear regression) and tree-based models, namely, decision trees, random forests, and popular gradient boosting models (LightGBM~\cite{lgbm}, XGBoost~\cite{xgboost}).
Furthermore, it supports representative featurizers: scalers, normalizers, and categorical feature encoders (\eg OneHotEncoder, LabelEncoder).




The algorithm
converts a model pipeline $M$ of a prediction query $P$ to an equivalent SQL statement.
First, it replaces each ML operator in $M$ with a corresponding SQL operator.
Linear models and scaling operators are converted to SQL using multiplication/addition/subtraction operators, while tree-based models and encoding operators are translated via \textsc{case} statements. For example, in case of decision trees, we do a depth-first search traversal of the tree nodes and create a nested \textsc{case} expression. For the {\sf TreeClassifier} in \autoref{fig:cross_ir_example}~(\ding{207}), the corresponding SQL expression is:

{\footnotesize
\begin{verbatim}
CASE WHEN F[0] > 60 THEN (
  CASE WHEN F[1] = 0 THEN 1 ELSE 0 END) ELSE (
  CASE WHEN F[2] = 1 THEN 1 ELSE 0 END) END    
\end{verbatim}
}


Once the per-operator conversion is done, \mltosql traverses the updated IR in topological order and each SQL operator is merged with its parents to produce a single SQL statement out of the IR.
\mltosql currently transforms the whole model pipeline $M$ to SQL or it fails (and $M$ is used instead). As part of our future work, we are investigating the benefit of a partial conversion.


Note that, as we discuss in \autoref{sec:opt:opt} and show in \autoref{sec:eval}, \mltosql should be used with care: it can be very beneficial in some cases but can be detrimental especially for complex models.

\vspace{1mm}
\mypar{\mltodnn} 
Unlike relational operators that are based on relational algebra and DNN operators that are based on tensor abstractions, traditional ML operators (\ie non-DNN ones, such as featurizers and linear/tree-based models) do not follow a similar computation abstraction.
Hence, despite its significance in the enterprise (as discussed in \autoref{sec:archi}), traditional ML has received less attention for optimization and hardware acceleration, as it is much harder to optimize arbitrary computation.

\sysname leverages recent work on Hummingbird~\cite{hummingbird} to translate traditional ML operators to equivalent DNNs that can be executed on highly efficient DNN engines like ONNX Runtime, PyTorch, and TensorFlow.
This is very important performance-wise, as DNN engines support out-of-the-box hardware acceleration through GPUs/FPGAs, as well as code generation~\cite{tvm}.
As we show in \autoref{sec:eval}, thanks to \mltodnn, we were able to evaluate prediction queries on SQL Server and Spark using GPUs for the trained pipelines of the queries. Note that given the overhead of moving data to the GPU, we will show that complex models benefit more from this optimization.

We are also investigating similar translations for the relational operators, so that we can hardware accelerate them without having to implement custom GPU kernels.

\subsection{Data-Driven Optimization Strategies}
\label{sec:opt:opt}



We now present novel optimization strategies that are employed by \sysname's optimizer to determine which optimization rules to apply on a given prediction query. We present an ML-informed rule-based and two ML-based strategies.





The input prediction query $P$ might include one or more \textsc{predict} operators, each invoking a trained pipeline (see \autoref{sec:archi}). 
\sysname triggers the optimization strategy on each sub-part of the IR that corresponds to a \textsc{predict} statement. In our running example, this is the part of the IR that corresponds to the trained pipeline $M$, depicted in the gray box in \autoref{fig:ravenflow}~(\ding{203}).

To perform data-driven decisions, we used the OpenML CC18 classification benchmark~\cite{openml} we described in \autoref{sec:over:motiv}.
We used $138$ models,\footnote{We excluded models that cannot be translated to ONNX, as well as a few multi-class classifications that we have not yet implemented support for.} which we executed using all combinations of our rules (both on CPU and GPU in the case of \mltodnn).
All OpenML evaluations were performed on Azure NC12s\_v2 instances, each with $12$ vCPUs, $224GB$ of RAM, and an NVIDIA Tesla P100 GPU. 

Our first observation is that \emph{our logical optimizations (\autoref{sec:opt}) are always beneficial}, as they reduce the required computation and data inputs when applicable, or leave the prediction query unchanged otherwise. Therefore, \sysname applies them in all cases. This intuition was confirmed in all OpenML runs.
The resulting speedup from these optimizations depends on the number of data predicates and unused columns, as we show in \autoref{sec:eval}.
We apply the predicate-based model pruning before model-projection pushdown, as the former can enable further application of the latter (see \autoref{sec:opt}). 

On the other hand, our OpenML runs showed that the impact of each logical-to-physical optimization (\autoref{sec:opt:optransform}) varies, as also shown in \autoref{sec:eval:micro}. 
Therefore, the optimization strategy has to choose between four 
evaluations for the trained pipeline: \mltosql (using the relational engine), \mltodnn (evaluating the resulting DNN on CPU or GPU), or none of them (using the ML runtime after applying only the cross-optimizations). Based on our runs, we excluded \mltodnn-on-CPU from the choices when a GPU is available, as it always gave worse or similar performance to \mltodnn-on-GPU.
Next we introduce different optimization strategies to pick between the above 
three  evaluations based on the model's characteristics.

\mypar{ML-informed rule-based strategy} Instead of hard-coding rules based on experience and magic numbers (like existing relational optimizers often do), this strategy follows a hybrid approach. It trains a decision tree based on our OpenML runs, uses it to find the $k$ most contributing features, and uses those to build a new much shallower decision tree that is turned to a rule. This also allows us to adapt to the specific hardware in hand. 
For each of the $138$ trained pipelines in the benchmark, we gathered $22$ statistics, including: \#inputs to the pipeline; \#inputs to model (after featurization); \#specific operators (e.g., one-hot encoders/OHEs); mean/max \#outputs of OHEs; \#trees, mean/max/stddev tree depth for tree-based models. As an example, using $k=3$ to strike a balance between simplicity and accuracy, the following rule was generated: \texttt{\small if \#features $>$$100$, apply \mltodnn; else if \#inputs $>$$12$ and mean tree depth $\le$$10$, apply \mltosql}.\footnote{Mean tree depth for linear/logistic regression is set to $0$.} The added benefit of this strategy is that no ML model needs to be invoked during optimization, which makes deployment in production simpler.



\mypar{Classification-based strategy}
In this strategy, we train an ML classifier using the OpenML dataset with the $138$ trained pipelines and the $22$ features mentioned in the rule-based strategy. The classifier predicts one of the following classes that correspond to the transformation to be applied: \mltosql, \mltodnn, none. We experimented with several scikit-learn~\cite{sklearn} classifiers (using default values), such as logistic regression, decision tree, random forest, and gradient boosting. Out of them, we use the random forest, as it gave best accuracy results.

\mypar{Regression-based strategy}
In this ML-based strategy, instead of building a classifier to directly pick the most promising transformation rule, we build a regression model that predicts the expected runtime after applying one of the \mltosql, \mltodnn, and no transformation variants. Therefore, the transformation becomes a feature in the regression model. The benefit of this approach is a 3-fold increase of the training set (one fold for every transformation option). At inference time, we perform three predictions, one for each transformation, and pick the option that gives the lowest runtime. We experimented with several scikit-learn regression models (e.g., linear regression, decision tree, random forest, XGB, adaptive boosting), and opted again for the decision tree due to best accuracy.

\begin{figure}[t]
	\centering\includegraphics[width=0.92\columnwidth]{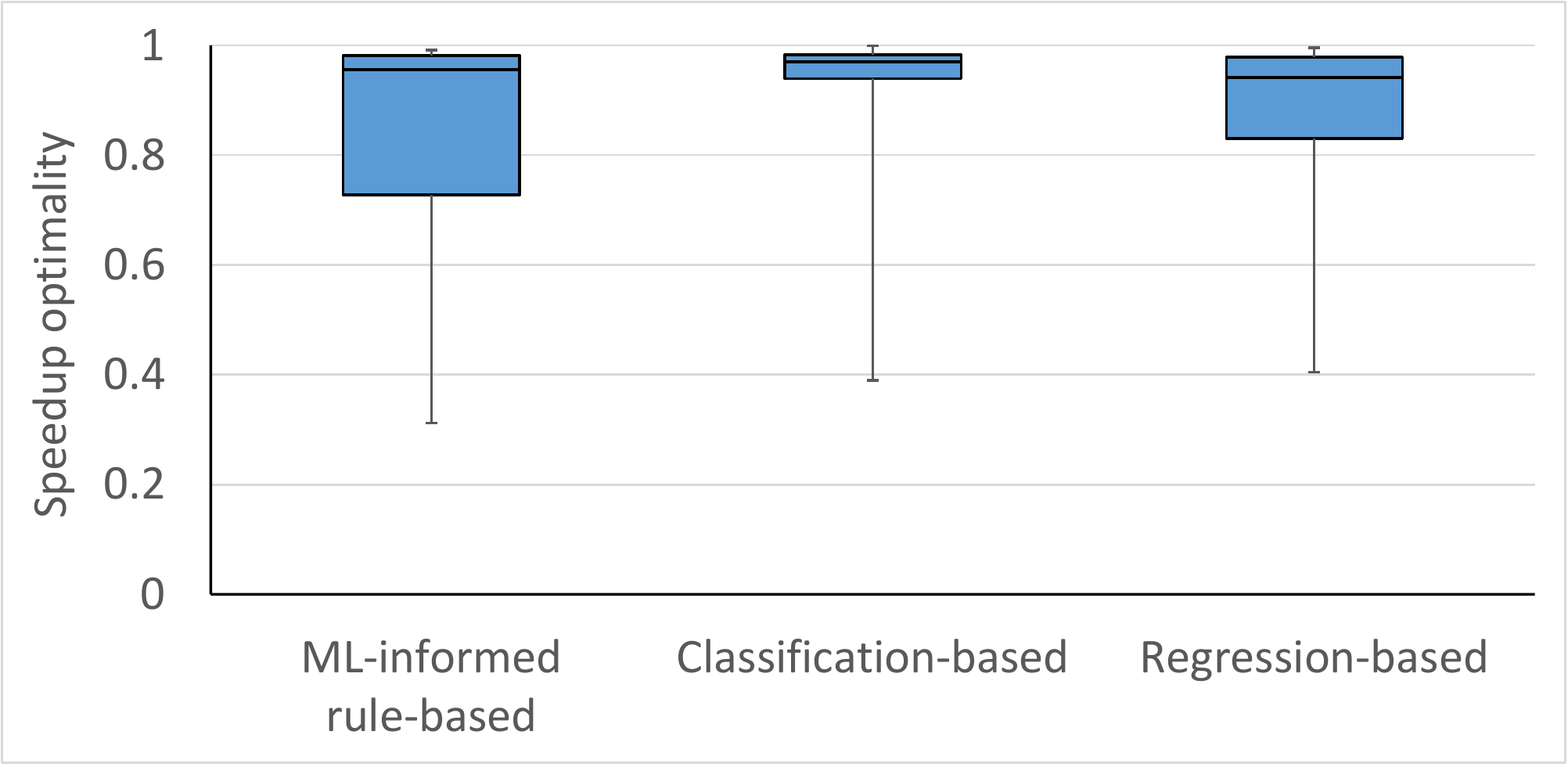}
 	\vspace{-3mm}
	\caption{Inference time speedup over optimal strategy achieved by different optimization strategies. Reported numbers are over $200$ stratified folds of the OpenML training dataset. Boxplots indicate 25$^{\textrm{th}}$, median, and 75$^{\textrm{th}}$ percentile; whiskers indicate min and max values.}
    \label{fig:optimizer-speedup} 
    \vspace{-6mm}
\end{figure}

\mypar{Strategy evaluation}
We now compare the performance of our strategies on the OpenML benchmark. Given our training set is relatively small and imbalanced ($25$ models perform best with \mltosql; $72$ with \mltodnn; $41$ with no transformation), we use stratified $5$-fold cross validation \srev{(with an 80/20 split between the training and test sets)} and repeat the process $40$ times for a total of $200$ runs. The rule-based had the lowest mean accuracy ($0.76$), while the ML-based ones both had $0.79$ (the regression-based had higher accuracy variance). \autoref{fig:optimizer-speedup} depicts the speedup achieved by each strategy (considering the total inference time to run all models of a run's test set) when compared to using the optimal transformation for each model. The median speedup of all three is comparable (classification-based is best with a speedup of $0.97$). However, classification-based has significantly lower variance, e.g., its $25^\textrm{th}$ percentile speedup is $0.94$ as opposed to $0.72$ and $0.83$ for the rule- and regression-based, respectively. In terms of training time, the $200$ runs took less than a minute for each strategy---users can go through this process once to finetune the strategy on their workload and hardware setup.


\vspace{1mm}
To sum up, \srev{\sysname first applies the logical optimizations in a strict order (predicate-based pruning before model-projection pushdown), and then it applies the logical-to-physical ones based on our data-driven strategies.} Among the ML-based strategies, the classification-based is preferable. The rule-based is a viable alternative when it is not desirable to invoke ML models during optimization.

\vspace*{-1mm}
\section{Implementation}
\label{sec:impl}

\setlength{\textfloatsep}{20pt}



\begin{figure}[t!]
	\includegraphics[clip, width=0.85\columnwidth]{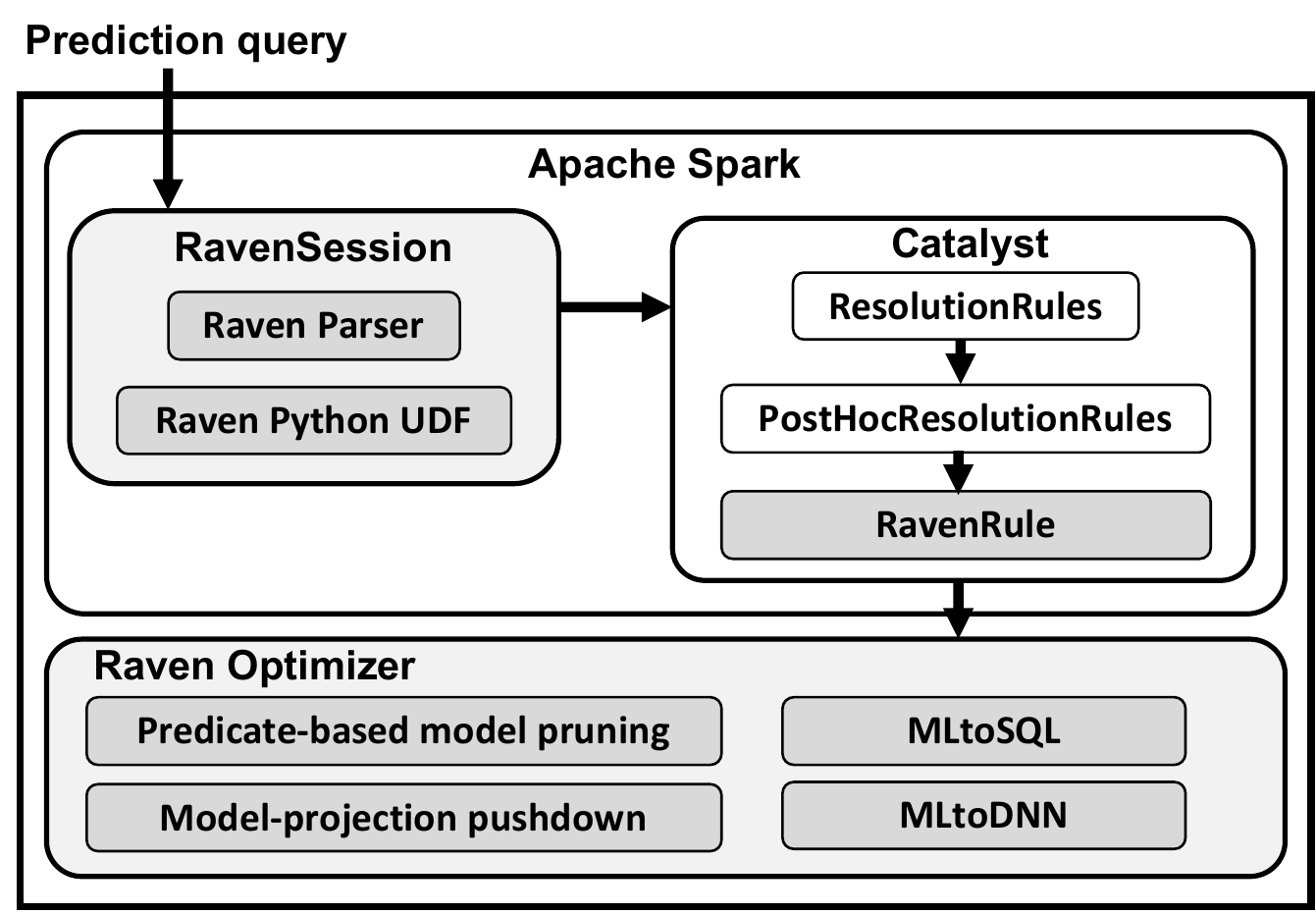}
  	\vspace{-3mm}
	\caption{\raven integration in Spark with the newly introduced components shown in gray.} 
    \label{fig:raven-implementation}
    \vspace{-6mm}
\end{figure}

We implemented \sysname as an extension to Apache Spark, depicted in 
\autoref{fig:raven-implementation} with the main \sysname components in gray.
In particular: 
(i)~we introduced the \textsc{predict} statement to allow SparkSQL users to express prediction queries, similar to~\cite{predictsql,big-query-ml,redshift-ml}; (ii)~we added a new rule to the Catalyst optimizer that internally invokes our \sysname optimizer, described in \autoref{sec:opt:opt}; and (iii)~we defined a Python UDF that allows invoking external ML runtimes (ONNX Runtime in the current implementation, as we support ONNX and ONNX-convertible models) to execute the prediction queries (on CPU and GPU).
In our integration, we made heavy use of Spark's extensibility framework: \sysname is provided as an external jar and no modification to Spark's source code is required. This was required to be able to deploy \sysname in any existing Spark installation and across \companyname's Spark offerings. \srev{The \textsc{predict} statement was made publicly available recently as part of our Azure Synapse Spark offering~\cite{sparkpredict}; we are actively working on productizing the optimizer, too.}

\sysname can be enabled programmatically by invoking the {\sf \small \raven\!Session} instead of the usual {\sf \small SparkSession}.\footnote{Note that the {\sf \scriptsize \raven\!Session} is actually a wrapper around the {\sf \scriptsize SparkSession}, thereby inheriting all {\sf \scriptsize SparkSession} functionalities.} 
Below we present the main components introduced, and then discuss how \sysname can generate queries to be executed in SQL Server.







\mypar{Adding the {\sc predict} statement to the parser}
We extended the {\sf \small ParserInterface} to allow users to submit SparkSQL queries containing {\sc predict} statements either as (i)~a UDF, e.g., {\sc select predict}(\texttt{model.onnx, *}) {\sc as predict from} \texttt{table}; or (ii)~as a table-valued function (TVF), e.g., {\sc select} \texttt{data.*, prediction.score} {\sc from predict(}\texttt{model = model.onnx, data = data)} {\sc with}\texttt{(score float)} {\sc as }\texttt{prediction}).
We internally rewrite \textsc{predict} TVFs into UDFs, so that we can have a single implementation for execution.
The \sysname parser detects \textsc{predict} statements and rewrites them to call the pre-defined \sysname Python UDF we introduced to invoke ML runtimes (see below). This allows us to hide the details of the \sysname UDF from the users and to trigger the \sysname optimizer when a \textsc{predict} statement is detected.


\mypar{Optimizing SparkSQL prediction queries}
We introduced a single {\sf \small PostHocResolutionRule} (``\sysname\!Rule'' in \autoref{fig:raven-implementation}) to trigger the \sysname (co-)optimizer when a \textsc{predict} statement is detected. This rule calls the \sysname optimizer as an external Python process. 
It also provides the optimizer with the predicates in the \textsc{where} clause of the query (for the predicate-based model pruning rule), along with the model~filename.

The \sysname optimizer uses the \texttt{onnxconverter\_common}  library~\cite{onnxconvert} to build a DAG of ONNX operators, containing all ML operators (see \autoref{sec:ir}) present in the prediction query. The data operators of the query are then added to the IR.
Once the full IR is built, the optimizer triggers all rules.
The output is the filename of the optimized model and a (possibly empty) set of columns (output of the model-project pushdown rule).
The rule then injects a new {\sf \small Projection} operator, and rewrites the \textsc{predict} with the new model~information.

\srev{To deal with the wide range of ONNX operators, we strive to share rule implementations across operators when possible. In particular, we share most of the rule implementation within linear models and within tree-based models (e.g., forests are just multiple trees). Featurizers are trickier and we tend to have custom code for each.}

\mypar{Running predictions with the Python Vectorized UDF}
We implemented the \sysname UDF using Spark's Python vectorized UDF~\cite{pandas-udf}, which allows to execute arbitrary Python code on a columnar view of input data, given as batches (of 10k tuples by default). Our UDF implementation performs the following: (i)~loads the model from HDFS;
(ii)~initializes and caches the model on a global variable to decrease model loading overheads in successive invocations on the same worker; 
(iii)~transforms the input columns to the right format; and (iv) invokes ONNX Runtime to perform the prediction.

\srev{Regarding the translation between relational data and model inputs (step (iii) above), most traditional ML models that we consider receive vectors as inputs (M$\times$1, with M being the batch dimension). Spark’s vectorized UDF automatically turns data to columnar Pandas dataframes, hence the translation is a 1-1 mapping. For models with M$\times$N inputs, we concatenate the M$\times$1 vectors within the UDF.}




\mypar{Transforming \sysname plans to SQL Server queries}
If specified by the user, \sysname can output a T-SQL version (SQL Server's SQL dialect)  of the optimized prediction query. 
Our goal with this design is to enable \sysname to output queries for other DBMSs too.

\vspace*{-1mm}
\section{Experimental Evaluation}
\label{sec:eval}

We now evaluate the benefits of \sysname on prediction queries over real-world datasets. To this end, we use \sysname to optimize the queries and execute them on Apache Spark and SQL Server (see \autoref{sec:impl}). We compare our results with several state-of-the-art systems for evaluating prediction queries, namely, SparkML, Spark invoking scikit-learn and ONNX Runtime, as well as MADlib (\autoref{sec:eval:macro}).
Then, through a set of micro-experiments, we study the impact of each optimization rule on  various model types (\autoref{sec:eval:micro}) and the benefit of GPU acceleration (\autoref{sec:eval:GPU}). We finally include a discussion on overheads, coverage, and accuracy.
%
%
Our key results are the following:
\begin{itemize}[noitemsep,topsep=3pt,parsep=0pt,partopsep=0pt,leftmargin=10pt]

\item On Spark, \sysname with all optimizations enabled delivers speedups of 1.4--13.1{\small$\times$}
against \sysname with no optimizations, and up to 48{\small$\times$} and 25.3{\small$\times$} speedups against SparkML and Spark with scikit-learn, respectively.

\item On SQL Server, \srev{\sysname plans provide speedups of 1.4--330{\small$\times$} over plans without \sysname optimizations}. Single-threaded \sysname gives speedups up to 108{\small$\times$} over MADlib for the supported queries.

\item For complex models, such as large gradient boosting models, \sysname can deliver significant speedups over GPUs: up to 8{\small$\times$} on Spark and 2.6{\small$\times$} on SQL Server. 

\item \rev{Based on the characteristics of each prediction query, different optimizations lead to optimal performance (as also shown in our optimization strategies of \autoref{sec:opt:logtophys})}.





\end{itemize}

\begin{table}[t!]
  \centering
  \caption{Summary of dataset statistics}
  \vspace{-3mm}
  \label{tab:dataset_stats}
  \setlength{\tabcolsep}{4pt} 
  \scalebox{0.9}{
  \begin{tabular}{cccc}
    \hline
    \multicolumn{1}{c}{Datasets} & 
    \multicolumn{1}{c}{\makecell[c]{\# of \\ tables }} &
    \multicolumn{1}{c}{\makecell[c]{\# of data inputs \\ (numeric \\ /categorical) }} & \multicolumn{1}{c}{\makecell[c]{\# of features \\  after encoding \\  (numeric/categ.)}}  \\
    \hline
    
    \multicolumn{1}{c}{\textbf{Credit Card}~\cite{credit-card}}&
    \multicolumn{1}{c}{1}&
    \multicolumn{1}{c}{\makecell{28  (28/0)}}&
    \multicolumn{1}{c}{\makecell{28  (28/0)}}\\
    
    \multicolumn{1}{c}{\textbf{Hospital}~\cite{hospdata}}&
    \multicolumn{1}{c}{1}&
    \multicolumn{1}{c}{\makecell{24  (9/15)}}&
    \multicolumn{1}{c}{\makecell{59  (9/50)}}\\
    
    \multicolumn{1}{c}{\textbf{Expedia}~\cite{hamlet}}&
    \multicolumn{1}{c}{3}&
    \multicolumn{1}{c}{\makecell{28  (8/20)}}&
    \multicolumn{1}{c}{\makecell{3965  (8/3957)}}\\
    
    \multicolumn{1}{c}{\textbf{Flights}~\cite{hamlet}}&
    \multicolumn{1}{c}{4}&
    \multicolumn{1}{c}{\makecell{37  (4/33)}}&
    \multicolumn{1}{c}{\makecell{6475  (4/6471)}}\\

    \hline
  \end{tabular}
  }
  \vspace{-3mm}
\end{table}

\mypar{Datasets} Throughout our experiments, we use four real-world datasets that have been widely used in data science tasks~\cite{credit-card,hamlet,hospdata}, summarized in \autoref{tab:dataset_stats}.
Some datasets are comprised of a single table while others of multiple ones, allowing us to vary the complexity of data operations from simple scans to multi-way joins.
The datasets include both numerical and categorical columns---the number of features per dataset (after encoding) ranges from 28 to 6,475 features.
In order to test prediction queries at scale \srev{and to match the dataset sizes we observe in production}, we replicate each dataset several folds, while making sure to not violate any primary-key constraints---for each experiment, we will specify the scale used.


\mypar{Trained pipelines}
We evaluate \sysname over four popular traditional ML model types~\cite{dsonds,kagglesurvey}, namely, logistic regression (\textbf{LR}), decision tree (\textbf{DT}), gradient boosting (\textbf{GB}), and random forest (\textbf{RF}). 
Each trained pipeline includes featurizers for numerical and categorical inputs: 
we normalize the former using standard scaling, and encode the latter using one-hot encoding~\cite{sklearn-scaler,sklearn-ohe}.
Trained pipelines all implement a binary classification task.\footnote{\sysname also supports several regression and multi-class tasks.}
Each pipeline is trained using scikit-learn over 80\% of the original (not scaled) datasets, and then converted into the ONNX format.
Each operator of the pipeline is trained using its default hyperparameters setting except when stated otherwise.


\begin{figure}[t!]
	\centering\includegraphics[width=1.0\columnwidth]{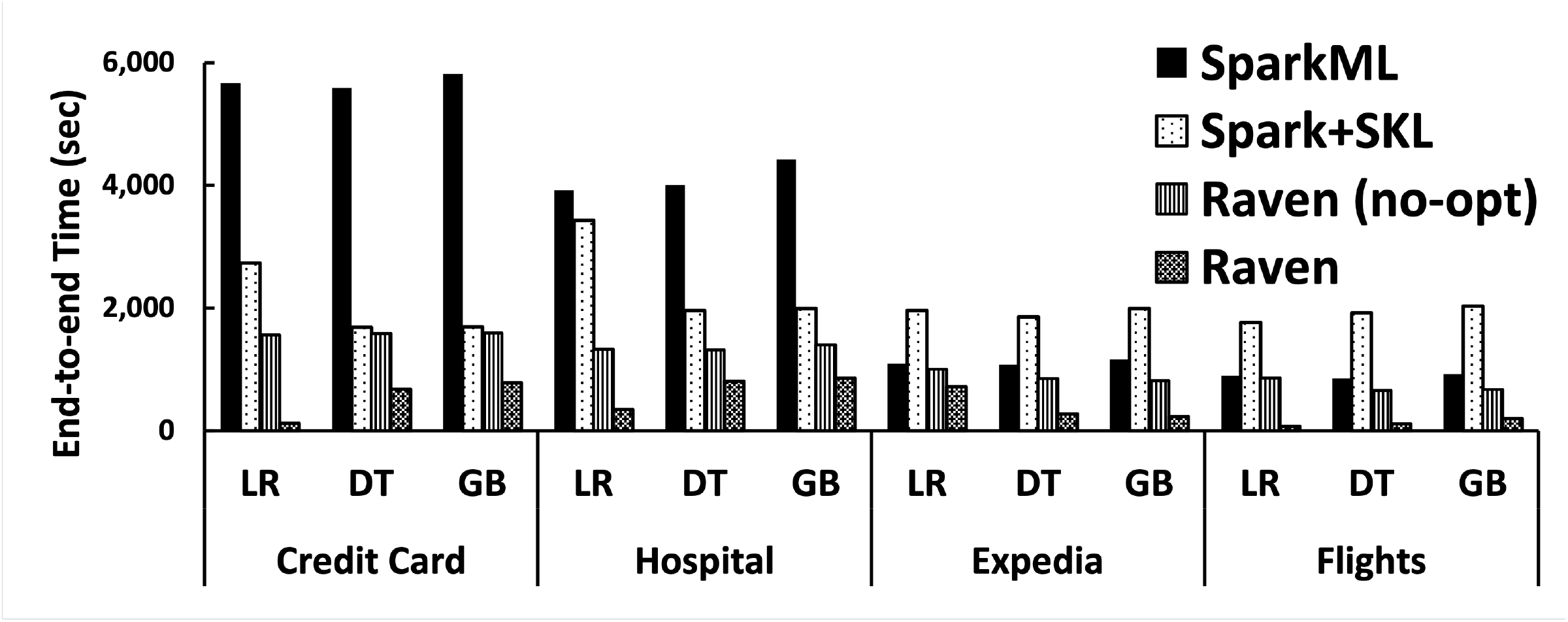}
	\vspace{-7mm}
	\caption{Prediction query runtime on Spark for different models and datasets, comparing \sysname with: SparkML, Spark with scikit-learn, and \sysname w/o  optimizations.} 
	\label{fig:macro}
	\vspace{-6mm}
\end{figure}

\mypar{Prediction queries} The prediction queries we use over single-table datasets (i.e., Credit Card and Hospital) involve a single table scan, whereas queries over Expedia and Flights involve a 3-way and a 4-way join, respectively. For the extra data predicates (e.g., \verb|asthma=1| in our running example), we add equality predicates in the {\sc where} clause of the queries.

\mypar{Reported metrics}
For each experiment we report the trimmed mean of the execution time of five runs, removing the lowest and highest runtimes.
For the Spark experiments, we report the total amount of time it takes to execute the prediction query and write the result to HDFS (with {\small \texttt{df.write.parquet(hdfs://...)}}) to avoid moving all results to the driver node. 
For SQL Server we add an aggregate operator on prediction results on the prediction queries.

\mypar{System setup}
For our Spark experiments, we set up a Spark v2.4.4 cluster on Microsoft Azure HDInsight~\cite{hdi}. It runs on YARN with 4 worker nodes and a driver node. Each worker has 56GB RAM and 8 cores, and the driver has 28GB RAM and 4 cores. We store data in Parquet~\cite{parquet} on Azure Block File System (ABFS). We leave the UDF batch size fixed at the default 10k rows per batch.
For SQL Server, unless otherwise specified, we used an Azure D32ds\_v4 VM instance, with 32 vCPUs, 128GB of RAM, and a 1.2TB SSD. We used the clustered columnstore index for all database tables.
For MADlib, we used version 1.17.0 on PostgreSQL 10.15.

Trained pipelines are authored using scikit-learn 0.21.3 (we use this version in inference comparisons, too). We use ONNX Runtime 1.2.0 and Hummingbird 0.2.1 (which internally uses PyTorch 1.7.1).

\vspace*{-1mm}
\subsection{End-to-end \sysname Evaluation}
\label{sec:eval:macro}

In this section, we evaluate \sysname against state-of-the-art frameworks, both distributed on Apache Spark (\autoref{sec:eval:spark}) and single-node on SQL Server (\autoref{sec:eval:SQL}). These experiments are on CPUs. We use the classification-based optimization strategy, as it gave us the best results (see \autoref{sec:opt:opt}). 


\subsubsection{\sysname on Spark}
\label{sec:eval:spark}

We compare our \sysname implementation on Spark (which uses ONNX Runtime as its ML runtime---see \autoref{sec:impl}) against \sysname with all optimizations disabled (\textsf{\small \sysname (no-opt)}) and the following widely used alternatives for distributed inference: (i)~SparkML; and (ii) a UDF-based approach similar to \textsf{\small \sysname (no-opt)} but with scikit-learn as its ML runtime (\systext{Spark+SKL}), to make sure \systext{\sysname (no-opt)} is competitive with state-of-the-art approaches.


\mypar{Datasets}
Credit Card is scaled to 1.6B rows; Hospital is scaled to 2B rows; Expedia is scaled to 500M rows; and Flights is scaled to 200M rows. 
We use a varying Hospital dataset size in the scalability experiment below.
\srev{We use (uncompressed) datasets between 200MB (1M rows Hospital dataset) and 800GB (Credit Card). This scaling aligns with our customer needs: most production datasets we see are between 1GB and 1TB.}

\mypar{Prediction queries}
For this experiment we pick three models: DT with max tree-depth of 8; LR with L$_1$-regularization and $\alpha$ (the regularization strength) set to 0.001; and GB with 20 estimators and max-depth of 3.
These models have, across all datasets, higher or similar AUC than the ones trained with default scikit-learn hyperparameters.
In \autoref{sec:eval:micro} we study \sysname's performance using different model parameters.
We implemented SparkML queries using PySpark with the same operators and settings as the scikit-learn pipelines.
Prediction queries in this experiment are run without data predicates.

\begin{figure}[t!]
	\centering\includegraphics[width=0.8\columnwidth]{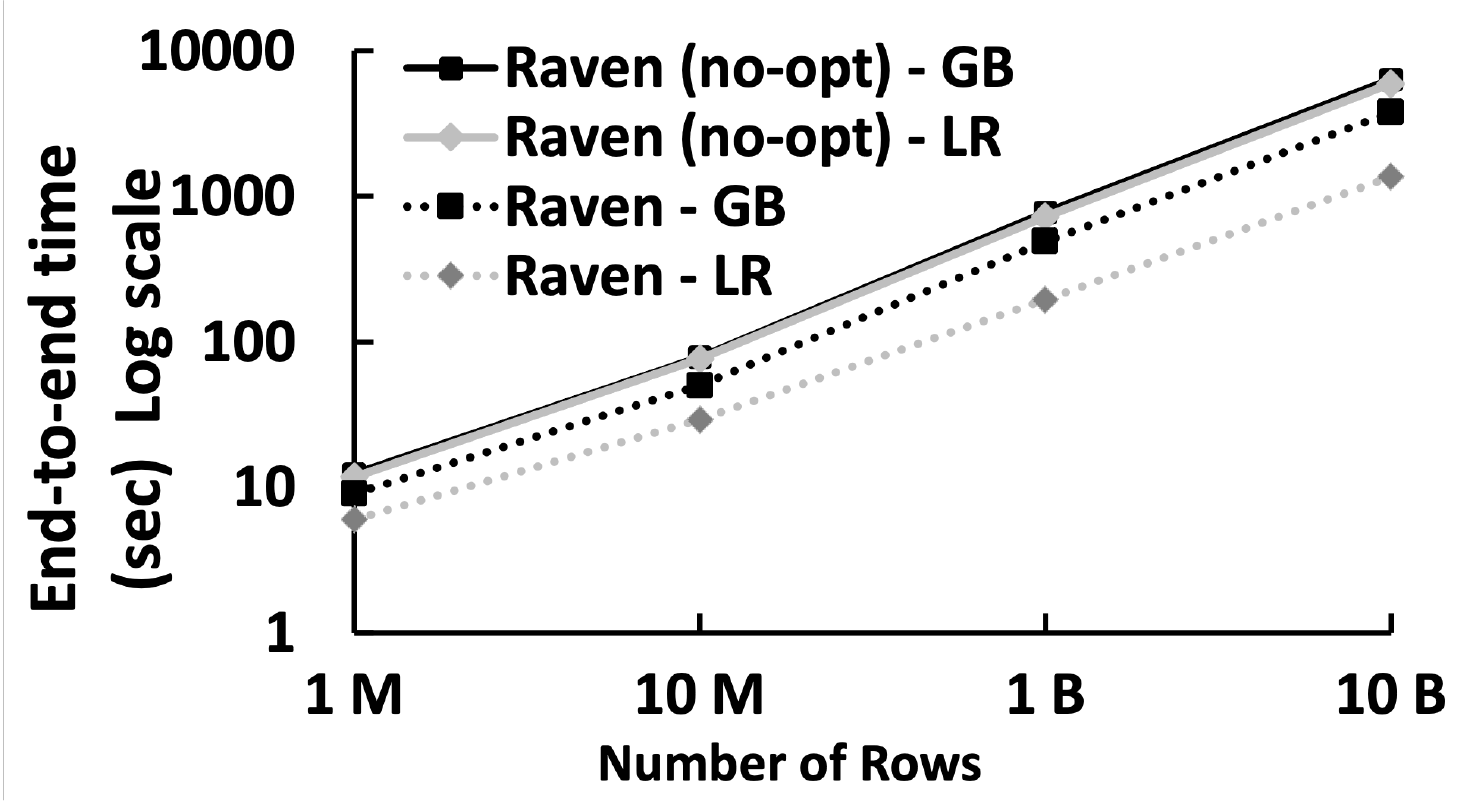}
	\vspace{-3mm}
	\caption{Performance of \sysname (with and without optimizations) for increasing Hospital dataset size.}
	\vspace{-5mm}
      \label{fig:diff-data-sz}  \vspaceafigure
\end{figure}

\mypar{Comparison with other systems}
The results for this experiment are reported in \autoref{fig:macro}. \sysname delivers 1.4--13.1{\small$\times$} speedup compared to \systext{\sysname (no-opt)}. Simpler models such as LR and DT benefit from both model-projection pushdown and \mltosql optimizations, avoiding reading unnecessary data by pruning unused data inputs and avoiding invoking ONNX Runtime (as the whole query is translated to SQL). This leads to a performance improvement of up to 13.1{\small$\times$}. 
Even when the optimizer selects no transformation rule, 
queries still benefit from \sysname's cross optimizations (i.e., model-projection pushdown---see details in \autoref{sec:eval:trees}). 
More interestingly, for Expedia and Flights (using 3-way, 4-way joins respectively), \sysname is able to push the projections on the unused inputs below joins, saving a lot of data movement and compute cost. Note that \mltodnn is not beneficial for any of those queries, and in fact it is never picked by the optimizer. As we will show later, more complex queries on GPUs make use of it.

Our results also show that \sysname is faster than the other frameworks, often by a large margin. 
SparkML is significantly slower than \sysname---between 1.5 and 48{\small$\times$}. 
Compared to \systext{Spark+SKL}, \sysname is 2.15--25.3{\small$\times$} faster. For single-table datasets, SparkML is slower than \systext{Spark+SKL} and \systext{\sysname (no-opt)}, but it is faster than \systext{Spark+SKL} and comparable to \systext{\sysname (no-opt)} when joins are involved.



\mypar{Data scalability} 
To study \sysname's scalability with increasing data size, we picked LR and GB from the above models (as \sysname triggers different optimizations for each of them) and compare \sysname with \systext{\sysname (no-opt)} for the hospital dataset with sizes of 1M to 10B tuples.
\autoref{fig:diff-data-sz} shows that \sysname consistently outperforms \systext{\sysname (no-opt)} for all models and dataset sizes: by 1.96--4.36{\small$\times$} for LR and by 1.37--1.67{\small$\times$} for GB. \sysname scales super-linearly to the data size, and this is likely due to the fact that for the small scales the cluster resources are not fully utilized.

\begin{figure}[t]
	\centering\includegraphics[width=1.0\columnwidth]{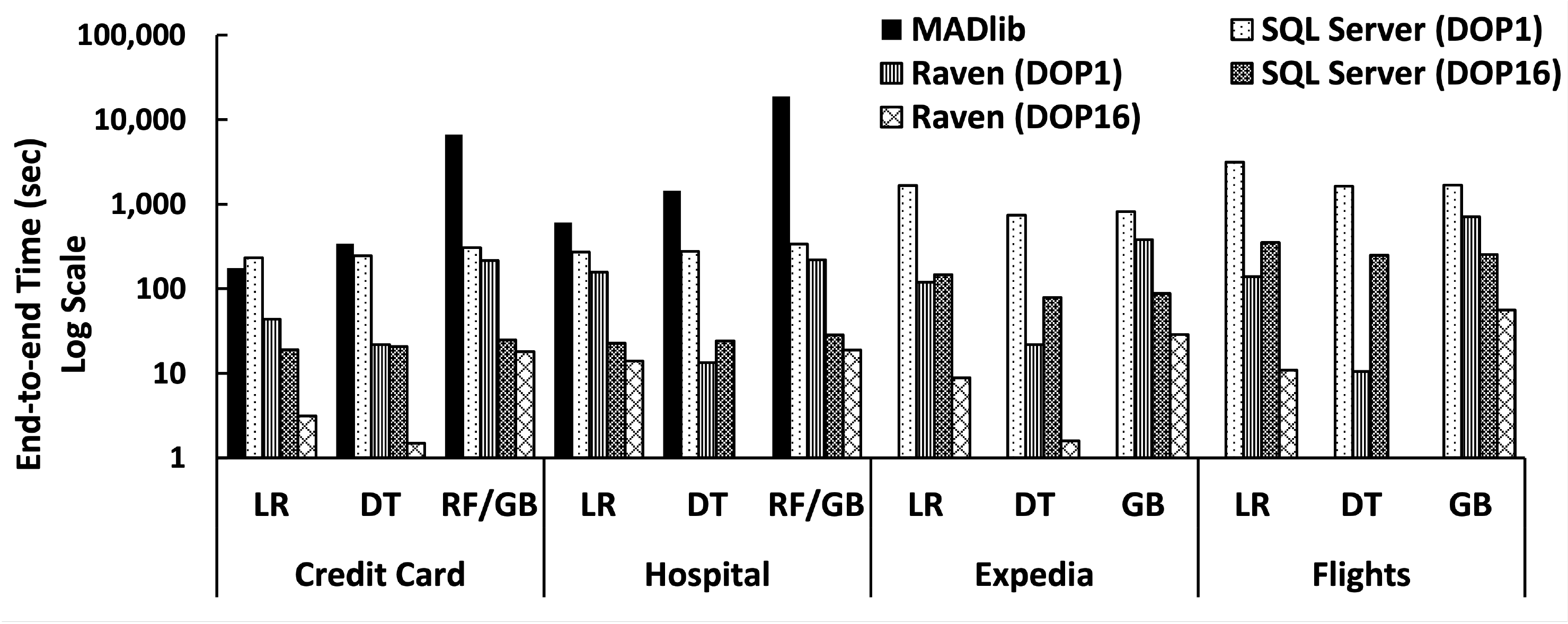}
	\vspace{-7mm}
	\caption{Prediction query runtime on SQL Server, both unoptimized and optimized by \sysname for degree-of-parallelism (DOP) 1 and 16, and on PostgreSQL with MADlib.} 
	\vspace{-7mm}
      \label{fig:macro-sql} 
\end{figure}

\begin{figure*}[t!]
    \minipage{0.32\textwidth}
    \centering
    \includegraphics[width=\linewidth]{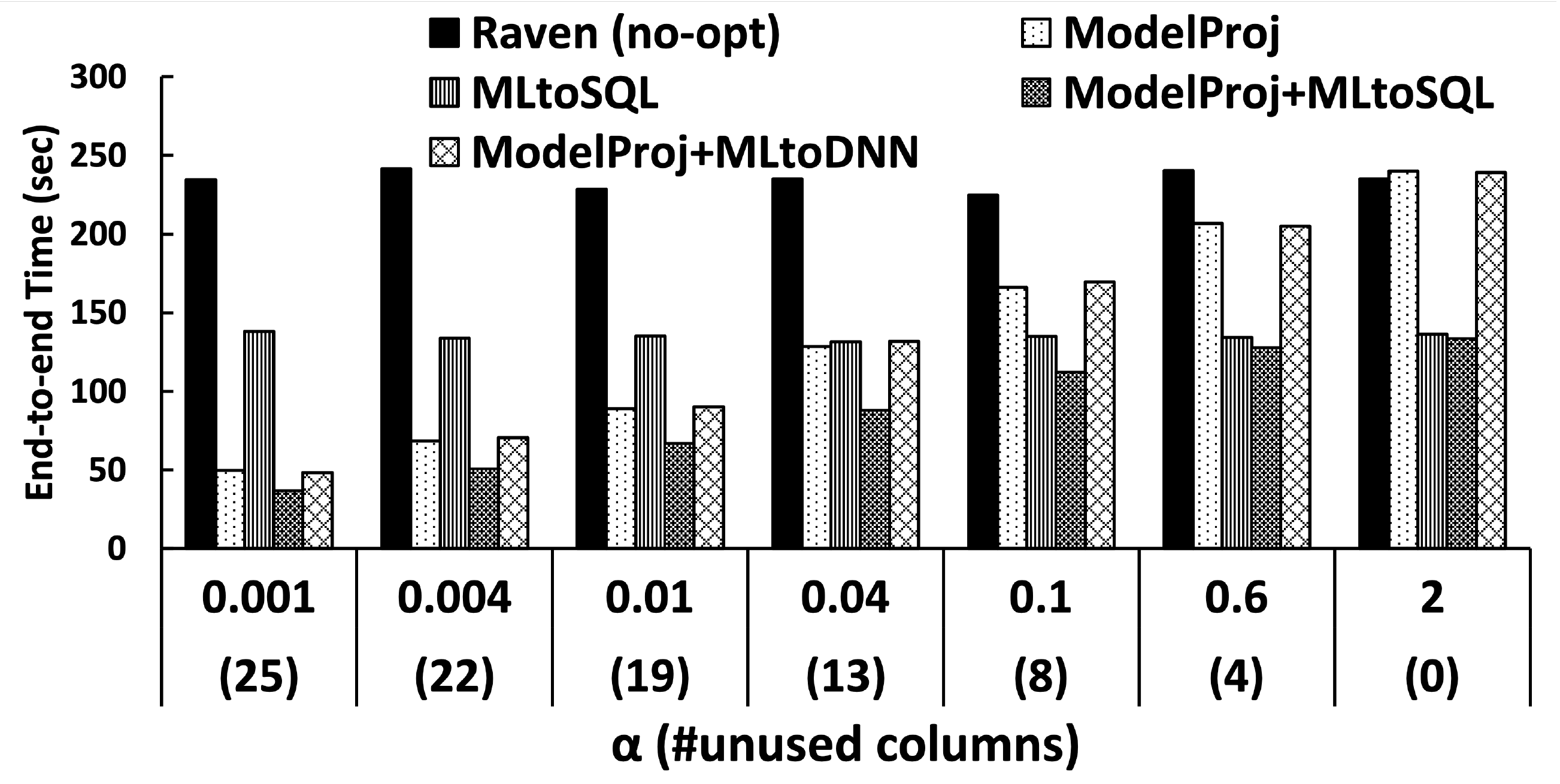}

    	\vspace{-2mm}
    	\caption{Impact of optimizations on linear models for the Credit Card dataset with varying regularization strengths $\alpha$.} 
          \label{fig:sparsity-cc}
    \endminipage\hfill
    \minipage{0.32\textwidth}
    \centering

        \includegraphics[width=\linewidth]{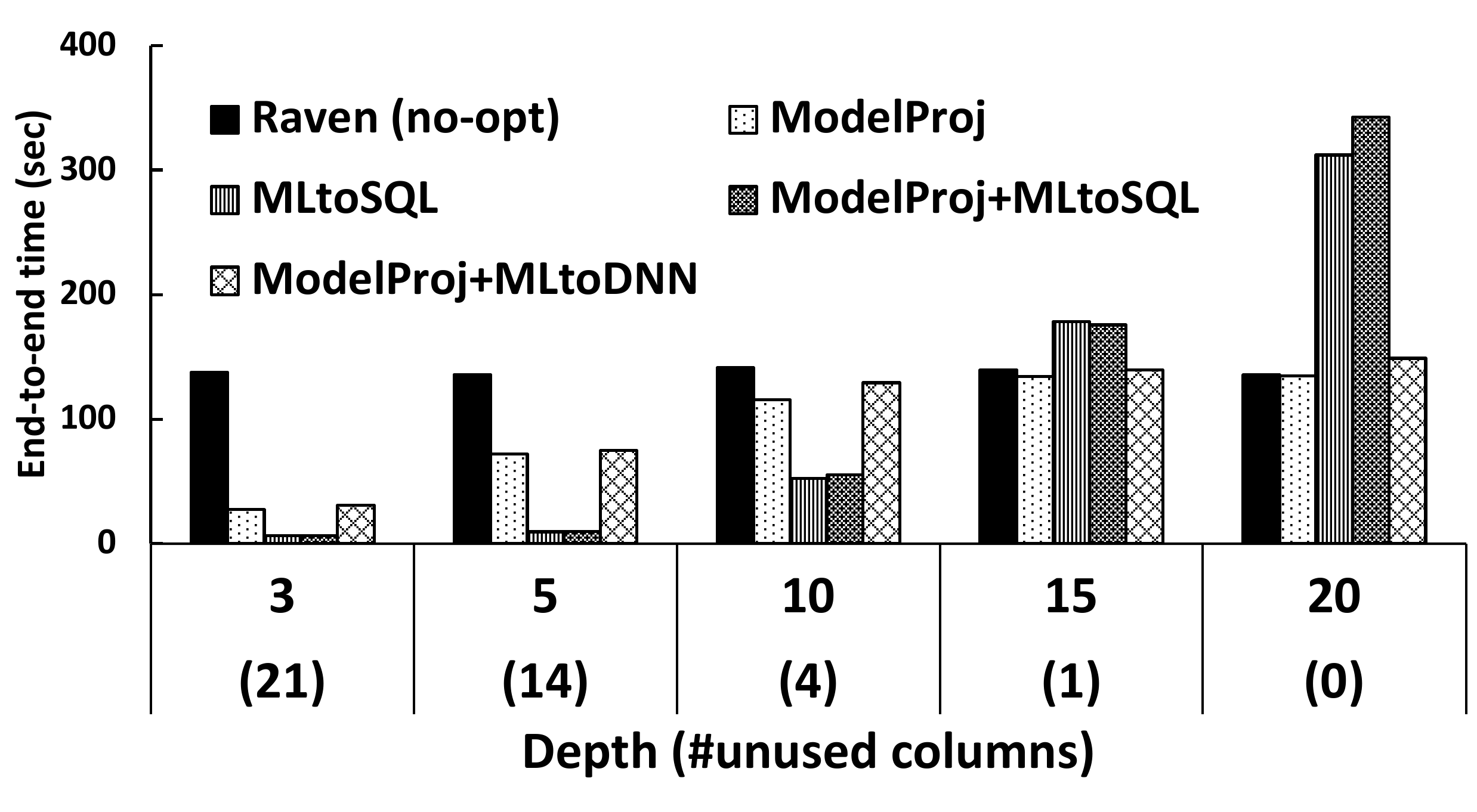}
    	\vspace{-5mm}
    	\caption{Impact of optimizations on decision trees for the Hospital dataset with varying tree depths. 
    	}
          \label{fig:complex-dt-hosp}
    \endminipage\hfill
    \minipage{0.32\textwidth}
    
    \centering
        \includegraphics[width=\linewidth]{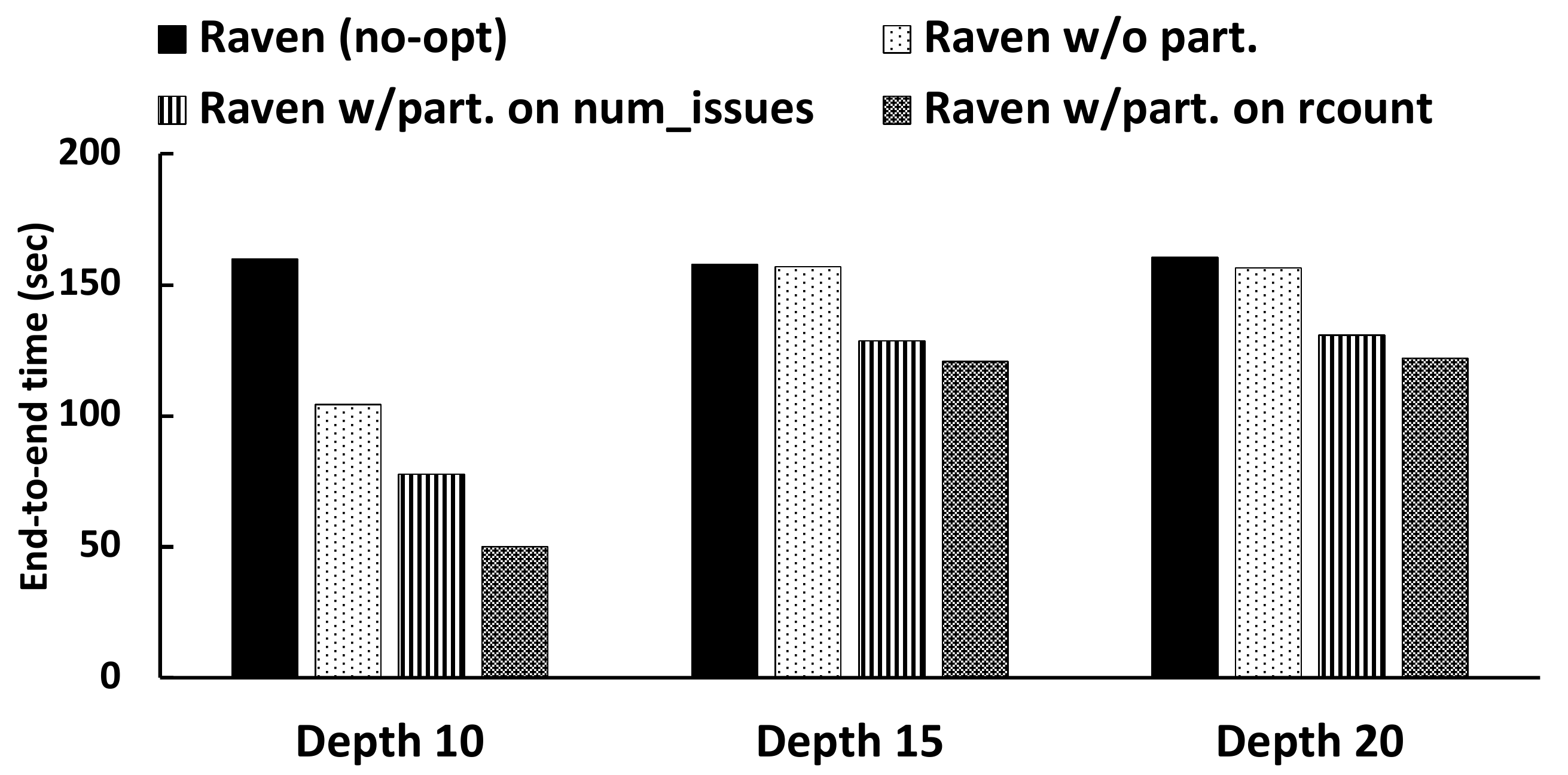}
        \vspace{-6mm} 
	\caption{Impact of data-induced optimizations on decision trees for the Hospital dataset with two partitioning schemes.}
      \label{fig:partitioned} 
    \endminipage
    \vspace{-5mm}
\end{figure*}

\subsubsection{\sysname on SQL Server}
\label{sec:eval:SQL}
Here we configure \sysname to output SQL Server queries (see \autoref{sec:impl}) and compare them with the original unoptimized queries on SQL Server. In both the optimized and non-optimized case we use SQL Server's \textsc{predict} statement that invokes ONNX Runtime to evaluate the model pipelines~\cite{predictsql,raven} \srev{along with all default SQL Server optimizations enabled.} We also compare against MADlib (on PostgreSQL), which is another well-known in-database ML engine~\cite{madlib}.
Since MADlib supports only single-threaded execution, for a fair comparison we report both single-threaded (with degree-of-parallelism 1 or DOP1) and multi-threaded (DOP16) execution for SQL Server (and thus \sysname). Unlike SQL Server, PostgreSQL does not support a columnstore layout.


\mypar{Datasets scale} We use all four datasets scaled to 100M rows. 

\mypar{Prediction queries} We use the same queries as in \autoref{sec:eval:spark} with the exception of the GB model in MADlib which we substitute with RF as this is the only tree ensemble model supported by MADlib. For the MADlib queries, we trained models using MADlib's ML training libraries with the same hyperparameters used for the original models.
Note that MADlib does not support pipelining of ML operations in most cases---instead we were forced to materialize the output of the featurization. For Expedia and Flights this led to surpassing the max number (1,600) of columns PostgreSQL allows in a table. Therefore, we skip these two datasets in the MADlib experiments.

\eat{
\mypar{Prediction queries} Prediction queries used in this section are \texttt{PREDICT} with single table scans (for Creditcard, Hospital), 3-way join (for Expedia), and 4-way join (for Flights). None of them have filters in their queries. Note that for all prediction queries, we added an aggregate operator on prediction results. For SQL Server and \sysname, we use SQL Server's in-process \texttt{PREDICT} UDF ~\cite{predictsql} in their prediction queries while
}

\mypar{Results} \autoref{fig:macro-sql} shows the results of our runs for \sysname on SQL Server, (un-optimized) SQL Server, and  MADlib. As discussed in \autoref{sec:eval:spark}, \sysname triggers model-projection pushdown for all models, but \mltosql only for LR and DT. Over SQL Server, it delivers speedups of 1.4--330$\times$. 
Specifically for DOP16, for LR and DT models, \sysname outperforms SQL Server by up to three orders of magnitude due to the translation to SQL and the removal of unused inputs, which are further pushed down below joins by SQL Server's optimizer.
Moreover, for the cases that the whole query gets translated to SQL, \sysname benefits from multi-threading more than the un-optimized query on SQL Server does.
\srev{The bigger speedup in the SQL Server experiments compared to the Spark ones is due to SQL Server’s more sophisticated optimizer. When \sysname turns a model to a SQL statement with ML-to-SQL, SQL Server can optimize it much more than Spark.}
Finally, single-threaded \sysname significantly outperforms MADlib by 3.9--108{\small$\times$}. This is to a large extent due to MADlib's materialization of intermediate featurization steps and its lack of \sysname optimizations.

\vspace*{-1mm}
\subsection{Micro-experiments}
\label{sec:eval:micro}

In this section, we explore the impact of each optimization rule (see \autoref{sec:opt}) on different model types. 

\subsubsection{Linear Models}
\label{sec:eval:linear}

We first study the benefits of \sysname's optimizations as we increase the sparsity of the linear model.
To this end, we use the Credit Card dataset with 200M rows on LR models with varying $L_1$-regularization strengths ($\alpha$ parameter). The lower the value of $\alpha$ the higher the regularization strength and the less the features with non-zero weights.

Our results are given in \autoref{fig:sparsity-cc}. The X-axis shows the $\alpha$ value and the resulting number of zero-weight inputs (out of the 28 total inputs). We report runtimes for different combination of rules (we use \systext{ModelProj} to denote model-projection pushdown) and for \systext{\sysname (no-opt)} with all optimizations disabled.

We see that the combination of \systext{ModelProj} and \mltosql is the best optimization for all LR model variants. Using only one of the two is not sufficient for realizing the full benefits of \sysname.
\systext{ModelProj} shows that as $\alpha$ increases, this optimization varies from only taking 20\% of the baseline run-time (with $\alpha$ = 0.001) to requiring slightly longer than the baseline (with $\alpha$ = 2), as the number of unused features decreases and thus the amount of data to be read increases. 
\mltosql alone is consistently about 60\% of the baseline run-time, as it avoids invoking the ML runtime.
The LR models in this experiments do not require too much compute, and therefore translation to DNN is not beneficial.


\subsubsection{Tree-based Models}
\label{sec:eval:trees}
In this section, we evaluate \sysname's optimizations over decision trees.
All experiments in this section were run on the Hospital dataset with 200M rows. 

\eat{
For our second experiment with \modelproj, we used 200 million rows of the hospital dataset and decision tree models. Similar to the previous experiment where we vary $\alpha$, in}

\mypar{Model complexity} In \autoref{fig:complex-dt-hosp} we show the impact of \sysname's optimizations as we increase DT's depth. The X-axis depicts the tree depth and (in parentheses) the number of columns, out of the 24, that are not used by the DT---the higher the depth the more inputs participate.
As we can see, model-projection pushdown is less effective as the tree depth increases, because less inputs are unused and can be pruned.

\mltosql provides a 21.7{\small$\times$} speedup for the depth-3 tree, but gets progressively less beneficial as tree depth increases, becoming a 2.3{\small$\times$} \textit{slowdown} for the final model. 
Note that the performance of \mltosql alone varies for tree models, whereas its impact on all LR models in \autoref{fig:sparsity-cc} was the same.
This is because, unlike \mltosql for LR, \mltosql for DT is automatically pruning unused features even without \systext{ModelProj}---it creates \textsc{case} statements (see \autoref{sec:opt:optransform}) only for paths with used inputs and the relational optimizer automatically projects out the rest. This is also the reason that \systext{ModelProj} does not provide additional benefit on top of \mltosql.
Finally, as with LR, \mltodnn is not a good option for small DTs on a CPU cluster. 

These results reinforce the importance of optimization strategies for picking the most efficient runtime, as shown in \autoref{sec:opt:opt}.

\mypar{Data predicates} In this experiment, we study the benefit of the predicate-based model pruning optimization by introducing an equality predicate in the {\sc where} clause of the prediction query for the Hospital dataset. We experimented with different DT models. For the decision tree with depth 20, which uses all 24 inputs, this optimization allowed to prune several tree branches based on the equality predicate. This led to an optimized trained pipeline with a shallower tree (i.e., depth 15) along with less data inputs (i.e., 23 inputs). \sysname saved $\sim$8\% of query time with predicate-based model pruning, and additional $\sim$12\% by removing 2 more data inputs with model-projection pushdown.

\begin{table}[t!]
  \centering
  \caption{Columns pruned by the data-induced optimization for three decision tree models on the Hospital dataset with different data partitioning schemes.} 
  \vspace{-3mm}
  \label{tab:partition_stats}
  \setlength{\tabcolsep}{4pt} 
  \scalebox{0.9}{
  \begin{tabular}{c|ccc}
    \hline
    \multicolumn{1}{c}  {\multirow{3}{*}{{\makecell[c]{Decision tree \\ depth}}}} &
    \multicolumn{3}{c}  {\makecell[c]{average \# of pruned columns}} \\
    
    \multicolumn{1}{c|} {\makecell[c]{}} &
    \multicolumn{1}{c|} {\makecell[c]{no \\ partitioning}} &
    \multicolumn{1}{c|} {\makecell[c]{partitioning on \\ \texttt{\small num\_issues}}} &
    \multicolumn{1}{c}  {\makecell[c]{partitioning on \\ \texttt{\small rcount}}} \\
    
    \hline 
    
    \multicolumn{1}{c|} {\makecell{10}}&
    \multicolumn{1}{c}  {\makecell{4}} &
    \multicolumn{1}{c}  {\makecell{8}} &
    \multicolumn{1}{c}  {\makecell{11}} \\

    \multicolumn{1}{c|} {\makecell{15}}&
    \multicolumn{1}{c}  {\makecell{0}} &
    \multicolumn{1}{c}  {\makecell{6}} &
    \multicolumn{1}{c}  {\makecell{5}} \\

    \multicolumn{1}{c|} {\makecell{20}}&
    \multicolumn{1}{c}  {\makecell{0}}&
    \multicolumn{1}{c}  {\makecell{6}} &
    \multicolumn{1}{c}  {\makecell{5}} \\
    
    \hline

  \end{tabular}
  }
  \vspace{-4mm}
\end{table}

\mypar{Data-induced optimizations} 
\rev{To assess the benefit of the data-induced optimizations (\autoref{sec:opt:datainduced}), we use the Hospital dataset, which we now partition in two different ways---on the `number of health issues' column (\texttt{\small num\_issues}) and the `readmission count' column (\texttt{\small rcount}), respectively. The former led to two partitions (whether or not there were health issues), while the latter led to six~partitions.}

\rev{\sysname uses the data-induced predicates to generate a different optimized (stripped-down) model for each partition.
We compare: (i)~\systext{\sysname (no-opt)}; (ii)~\sysname w/o partitioning, which is the ``best-of'' our prior optimizations (when data is not partitioned, the data-induced optimization is not applied); and (iii)~\sysname on the two partitioned versions of the dataset with all optimizations enabled.
\autoref{fig:partitioned} shows the impact of the data-induced optimizations on the end-to-end time of scoring 200M rows of the Hospital dataset for decision trees of varying depth, while \autoref{tab:partition_stats} shows the number of columns that were pruned on average across the optimized models.
For decision trees with depth 15 and 20, \sysname with the data-induced optimization improves the performance by $\sim$20\% due to the additional tree pruning based on the induced predicates (see \autoref{tab:partition_stats}) whereas \sysname w/o partitioning is similar to \systext{\sysname (no-opt)}. For depth 10, \sysname with partitioning provides 2.1--3.2{\small$\times$} and 1.3--2.1{\small$\times$} performance improvements over \systext{\sysname (no-opt)} and \sysname w/o partitioning, respectively, due to pruning more tree branches along with the \mltosql transformation. Although the exact benefit depends on the partitioning column, it is beneficial in both cases.}

\begin{figure}[t]
	\centering
		\includegraphics[width=0.8\columnwidth]{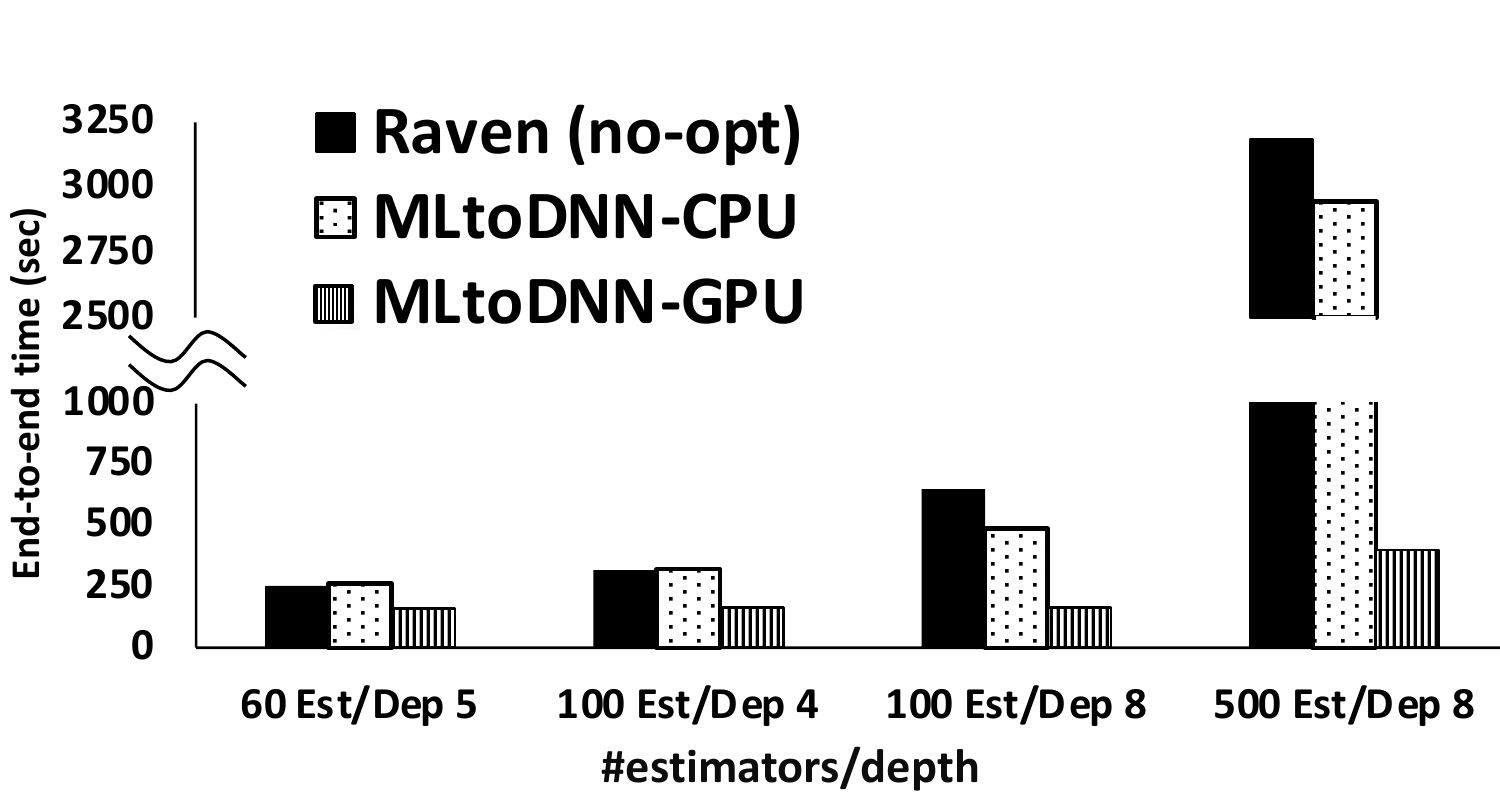}
	\vspace{-3mm}
	\caption{Impact of \mltodnn rule over CPU and GPU on complex gradient boosting models for the Hospital dataset.}
	\vspace{-6mm}
      \label{fig:gpu} 
\end{figure}


\vspace{-1mm}
\subsection{GPU acceleration of complex models}\label{sec:eval:GPU} 

Although we showed that \mltodnn is not beneficial for simple models, using GPUs can greatly decrease inference time for more complicated ones, which are not uncommon as we saw in \autoref{sec:archi}. 
For this experiment, we used a separate Spark cluster with one driver node and three worker nodes, each with 6 CPUs, 56 GB memory, and NVIDIA Tesla K80 GPUs. We purposely picked this setup to have an hourly cost  close to the CPU Spark cluster we used.

Unlike \autoref{sec:eval:macro} where we used 20 estimators with max tree-depth 3, here we use 60--500 estimators with max depth 4--8. For these models, model-projection pushdown is not beneficial as all inputs are used, and \mltosql is detrimental as it generates too complex SQL expressions.

\autoref{fig:gpu} shows our results to compare \sysname without optimizations with \sysname that triggers \mltodnn.
The resulting plan includes a PyTorch DNN (instead of the original GB model) that we execute over both CPU and GPU (the remainder of the plan is executed on CPU) for the Hospital dataset.
We observe speedups of 1.56--7.96{\small$\times$} on GPU. Note the break in the graph: the 500 estimators/depth 8 model took 3,184 seconds on CPU without optimizations and only 400 seconds when run on GPU. In general, the more complicated the model, the bigger the speedups on GPU.
Depending on the complexity of the model, \mltodnn can be beneficial on CPU, too:
the first two models have a minimal slowdown due to overhead, but the rightmost models have sizable speedups of 1.08--1.33{\small$\times$}~on~CPU.

We performed the same experiments for SQL Server on an Azure NC6s\_v3 instance with a Tesla V100-PCIE-16GB GPU. 
\mltodnn produced speedups of 2.3--2.6{\small$\times$} across the four models. For these runs we use SQL Server's Python extensibility framework~\cite{spexternalsql}---using a tighter integration~\cite{predictsql} should further increase benefits.

\subsection{Discussion}
\mypar{Optimization overheads in Spark}\label{sec:perf:overhead}
Our (un-optimized) inference queries on Spark showed an overhead of
2--4 seconds for cold runs and $\sim$0.1 second on warm runs, related to invoking the ML runtime through the Python UDF and loading the model from HDFS. Although optimization time varies based on model type and dataset, model-projection pushdown
takes 1--5 seconds, \mltosql takes 3--5 seconds, and \mltodnn takes 0.1--0.5 seconds on warm runs. Among our experimental prediction queries, the complex ones with hundreds of nodes in the IR tend to take up to 5 seconds for optimizations.
On our four-node Spark cluster, we found that $\sim$1M rows were a sufficient quantity across all of our datasets to mask the optimization overheads (see \autoref{fig:diff-data-sz}). 
\srev{We are investigating optimizations that could further reduce these overheads.
Moreover, some of our optimizations could be performed offline (saving the optimized model/plan)---this way \sysname can be beneficial for any dataset size.}

\srev{The main overheads introduced by the {\sc predict} statement are from (i)~the model loading from disk, (ii)~the startup time for the UDF (i.e., the Python runtime startup cost), and (iii)~the data conversion from Spark internal row representation to the format used by the models.
For (i), we are already exploring how to cache/reuse models within Spark. 
For (ii), an alternative is to implement the UDF directly in Scala. However, this will force us to call the ML runtime using JNI, which will introduce other overheads.
For (iii), when calling a {\sc predict} UDF, Spark converts its row format to Arrow and then to Pandas dataframes. Most overhead comes from the conversion to Arrow, while the Arrow-to-Pandas conversion is zero-copy.}

\mypar{Coverage} To study the coverage that our IR and our optimizations provide in terms of ML operators, we trained 508 scikit-learn pipelines from the OpenML CC-18 benchmark~\cite{openml}, which we converted to ONNX. We found that all operators can be expressed in our IR. Regarding the applicability of \sysname's optimizations, model-projection pushdown can be performed on all pipelines, \mltosql lacks support for just $4$ operators across all pipelines, and \mltodnn covers 88\% of all pipelines.

\mypar{Prediction accuracy}
To investigate whether our operator transformations (\autoref{sec:opt:optransform}) introduce rounding errors, 
we compared the result of the optimized \sysname plan with the unoptimized one for all our models in Spark and SQL Server.
In the 30 models that we examined,
\mltosql led to rounding errors in $0.006$--$0.3$\% of the predictions, whereas \mltodnn in less than $0.8$\%~\cite{hummingbird}. 
%
%
Such negligible rounding errors are considered acceptable for ML converters~\cite{onnxconvert}.

\section{Related Work}
\label{sec:relwork}

The history of integrating ML algorithms in databases is rather long. In the early 2000's, SQL Server shipped with data mining operators~\cite{sqlserver-data-mining}. Later in the 2010's, MADlib~\cite{madlib} suggested using UDAs/UDFs for embedding ML into the database. Apache Spark’s MLlib~\cite{sparkmllib}, Apache Mahout Samsara~\cite{samsara} and others~\cite{mli,keystone,bismark,learning-joins,db4ml,in-db-ml,DBLP:journals/pvldb/ZhangMJKKKVK21} can be seen as a continuation of this effort. A broad overview of in-DB ML techniques can be found in~\cite{db-in-ml-tutorial}.
All these works, however, focus mostly on the training aspect of ML that is heavy on iterations. Moreover, ML components are handled as black boxes, thereby limiting optimization opportunities.

More recently, Amazon, Google, and Microsoft have added support for in-DB ML inference~\cite{predictsql,raven,big-query-ml,redshift-ml}, also focusing on enterprise capabilities of managed databases~\cite{cloudy}.
\sysname follows the architectural trend of collocating data engines with ML runtimes for inference, and takes it a step further by co-optimizing relational and ML operators and using the most efficient runtime for each part of the prediction query. \srev{As part of our broader effort to make it easier to author inference queries using any model, we have made the {\sc predict} statement available in several of our engines, including Azure Synapse SQL and Spark, and Azure SQL Edge~\cite{predictsql,sparkpredict,predictedge}.}


\srev{The \sysname optimizer we presented in this paper realizes and extends the vision that we first introduced in~\cite{raven} in the following ways: we (i)~built an end-to-end optimizer for applying the logical optimizations, (ii)~added the data-induced optimizations, (iii)~introduced data-driven strategies for the logical-to-physical optimizations, (iv)~provided several production evidence data, the OpenML study, and (v)~an extended experimental evaluation. Our initial prototype of~\cite{raven} could manually apply just two logical optimizations over single-operator linear regression and decision tree models, but real models include at least tens and up to thousands of operators. The optimizer we presented here covers all OpenML pipelines for the logical optimizations and 88\% for the physical ones---our initial prototype could handle none of these models.}

\rev{SystemML explores cross-optimizations between relational and ML operators, but, unlike \sysname, focuses on the training part.
Probabilistic predicates~\cite{probabilistic-predicates} is a method to learn predicates and push them below trained pipelines for optimizing inference---such techniques could be integrated with \sysname, too.
More recently,~\cite{kang2020jointly} co-optimizes featurization and deep learning models.
Thanks to its whitebox approach, \sysname is able to optimize arbitrary ML pipelines containing featurizers, trees, and algebraic models.}

\rev{Tensorflow's \texttt{\small tf.data}~\cite{tf-data} provides optimizations for data pipelines feeding Tensorflow models, but no cross-optimizations between data and ML are explored.
Teradata's recent Collaborative Optimizer~\cite{DBLP:journals/pvldb/EltabakhSAANHCZ21} pushes projections and predicates over analytical functions. \sysname's cross-optimizations are related but go further by specifically targeting prediction functions over ML models, which they can optimize, e.g., by pruning the model or compiling it into SQL statements.
Froid~\cite{froid} compiles a limited class of UDFs into SQL.
Our data-induced optimizations have similarities with the data-induced predicates used to optimize relational queries~\cite{dips19}.
SQL4ML~\cite{sql4ml} translates ML operators implemented in SQL into Tensorflow components such that training can be efficiently executed in Tensorflow.}

\rev{Recent works have used a common IR to express both relational and ML operators~\cite{mlinspect,lara,laradb}, focusing on ML training and on supporting relational and linear algebra. Similarly, another line of work~\cite{ml2sql,systemml,simsql,rafiki,aida}
integrates tensor types and linear algebra routines within a data processing system for efficiently handling in-engine SQL and ML workloads.
Although these are useful for generalized linear models and DNNs that rely heavily on linear algebra, they cannot target tree-based methods and featurization operators that are among the most used ML operators (see \autoref{sec:archi}).
Weld~\cite{weld} introduces a new set of abstractions and runtime for speeding up Python programs containing Pandas and scikit-learn models. It offers an IR targeting physical optimizations that could be used after \sysname's logical optimizations.}








The wide deployment of ML models has also drawn the attention of the systems community to focus on efficient inference, beyond the training of ML models.
Popular approaches for model inference~\cite{fromtheedgetothecloud} include containerized~\cite{clipper} and in-application~\cite{mlnet} execution.
ML frameworks built from the ground up for inference workloads are also emerging~\cite{ort,tvm}. 
Pretzel and Willump are optimizers specifically built for inference workloads~\cite{pretzel,willump}, but, unlike \sysname, they are not capable of holistically optimizing inference queries.

\eat{
\color{lightgray}
Several previous works have proposed to run machine learning into the RDBMS~\cite{madlib,simsql2,systemml,levelheaded,doppiodb}. 
Interesting enough, these works mostly focus on training, whereas the prime focus of \raven is inference of already trained machine learning models. 
Other popular approaches for model inference~\cite{fromtheedgetothecloud} are containerized~\cite{clipper} and in-application~\cite{mlnet} execution.
As we argued in the introduction, model inference has interesting algebraic characteristics which makes it a likely (easier) target of integration with database query optimizer and runtime.
Cross-optimization of relational and linear algebra operators is recently becoming a hot topic~\cite{laradb,lara}, whereas specific optimizers~\cite{pretzel,willump} and runtimes~\cite{onnx,tvm} for model inferences are starting to emerge as well.
Our goal with \raven is to bridge the gap between the two worlds: we propose an optimizer able to execute both runtime-specific and cross-IR optimizations in an end-to-end~fashion.

New paper (Stanford that KK found and MI reviewed) similar to raven~\cite{kang2020jointly}, it is the same premise as \raven but has nothing to do with the \raven techniques.

In~\cite{DBLP:journals/corr/abs-1906-01974}, "an optimizer for ML inference that introduces
two statistically-motivated optimizations targeting ML applications whose performance bottleneck is feature
computation".
}

\color{black}

\section{Conclusion}
\label{sec:future}

\rev{Efficient execution of prediction queries, which involve data processing operators and trained models for inference, is key for the success of machine learning in the enterprise.
In this paper, we exploit the recent trend of collocating data engines with ML accelerators and attempt to break the silos between relational and ML operators.
The result is \sysname: a production-ready system that is able to not only push information from relational operators into ML models and vice-versa to holistically optimize the queries, but it can also pick the right target runtime (e.g., SQL engine or ML/DNN framework) for each operator and exploit hardware accelerators, based on learned optimization strategies. This translates to orders of magnitude improvements over state-of-the-art solutions on both Apache Spark and SQL Server.
\sysname's extensible architecture enables to easily add new optimizations as part of our future work, e.g., to push relational operators to the ML accelerators.}


\eat{
{\color{gray}
We presented \sysname, a system we are building  to
perform in-DB ML inference. \sysname performs static analysis of Python ML pipelines and SQL queries, that are captured in a unified IR. This enables us to apply novel cross-optimizations, yielding performance gains of up to $24\times$. The target execution environment for this optimized IR is a deeply integrated version of SQL Server and ONNX Runtime, which alone provides up to $5.5\times$ performance gains over
standalone ONNX Runtime execution. This is only the beginning of a long journey to incorporate ML scoring as a foundational extension of relational algebra, and an integral part of SQL query optimizers and runtimes.
}
}


\begin{acks}
\end{acks}

\srev{We would like to thank the following people that contributed to this work through their insightful feedback and collaboration: Carlo Curino, Nellie Gustafsson, Andreas Mueller, Ivan Popivanov, Raghu Ramakrishnan, Markus Weimer, Doris Xin, Yuan Yu, Yiwen Zhu.}


\balance

\bibliographystyle{ACM-Reference-Format}
\bibliography{references}

\end{document}